\documentclass[preprints,article,accept,oneauthor,pdftex,a4paper]{Definitions/mdpi}
\firstpage{1}
\pubvolume{xx}
\issuenum{1}
\articlenumber{5}
\pubyear{2019}
\copyrightyear{2019}
\history{Received: date; Accepted: date; Published: date}

\pdfoutput=1

\usepackage{amsmath}
\usepackage{amsfonts}
\usepackage{amssymb}
\usepackage{accents}
\newcommand{\dd}{\mathrm{d}}
\newcommand{\DD}{\mathrm{D}}
\newcommand{\lc}[1]{\accentset{\circ}{#1}}

\Title{Metric-affine Geometries With Spherical Symmetry~$^{\dagger}$}

\Author{Manuel Hohmann $^1$}

\AuthorNames{Manuel Hohmann}

\address[1]{%
$^{1}$ \quad Laboratory of Theoretical Physics, Institute of Physics, University of Tartu, W. Ostwaldi 1, 50411 Tartu, Estonia; manuel.hohmann@ut.ee}

\corres{Correspondence: manuel.hohmann@ut.ee}

\firstnote{This paper is based on the talk at the 2nd International Conference on Symmetry, Benasque, Spain, 1-7 September 2019.}

\abstract{We provide a comprehensive overview of metric-affine geometries with spherical symmetry, which may be used in order to solve the field equations for generic gravity theories which employ these geometries as their field variables. We discuss the most general class of such geometries, which we display both in the metric-Palatini formulation and in the tetrad / spin connection formulation, and show its characteristic properties: torsion, curvature and nonmetricity. We then use these properties to derive a classification of all possible subclasses of spherically symmetric metric-affine geometries, depending on which of the aforementioned quantities are vanishing or non-vanishing. We discuss both the cases of the pure rotation group \(\mathrm{SO}(3)\), which has been previously studied in the literature, and extend these previous results to the full orthogonal group \(\mathrm{O}(3)\), which also includes reflections. As an example for a potential physical application of the results we present here, we study circular orbits arising from autoparallel motion. Finally, we mention how these results can be extended to cosmological symmetry.}


\begin{document}
\section{Introduction}\label{sec:intro}
By its geometric nature, the description of gravity within the theory of general relativity stands out from all other field theories. The quest for a unified field theory, together with tensions posed on general relativity both by cosmological observations and its consistence with quantum theory, have therefore led to the development of a plethora of alternative gravity theories~\cite{Capozziello:2011et}. The fact that all other forces of nature are modeled by gauge theories, so that the fields mediating these interactions are described by connections on principal bundles, motivates a similar approach to gravity, thus introducing a connection as a fundamental field to mediate the gravitational interaction. Further taking motivation from the idea that gravity is linked to the geometry of spacetime itself, the most straightforward choice is to consider a connection in the frame bundle of spacetime, i.e., an affine connection. Finally, taking into account that observations suggest to use a Lorentzian metric to describe the dynamics of fields and particles on spacetime, one arrives at the notion of metric-affine geometry, and hence the class of metric-affine theories of gravity~\cite{Hehl:1994ue}.

Particular subclasses of metric-affine geometries appear in various theories of gravity. Einstein-Cartan gravity~\cite{Kibble:1961ba,Sciama:1962}, and the more general class of Poincaré gauge theories~\cite{Blagojevic:2002du,Blagojevic:2013xpa}, make use of a metric-compatible connection. In general relativity, this is further specialized to the unique metric-compatible, torsion-free connection, which is the Levi-Civita connection. The latter constitutes one corner of the so-called ``geometric trinity of gravity''~\cite{BeltranJimenez:2019tjy}, which in addition comprises of the equivalent formulations of general relativity in terms of metric teleparallel~\cite{Aldrovandi:2013wha,Maluf:2013gaa} and symmetric teleparallel~\cite{Nester:1998mp} geometries. Combining the latter two, one arrives at a general teleparallel geometry, featuring both torsion and nonmetricity~\cite{Jimenez:2019ghw}. Numerous modified gravity theories based on these geometries have been studied.

An important task in the study of gravity theories is determining exact solutions to their field equations. This task is often simplified by considering solutions with spacetime symmetries, i.e., invariance under the action of a symmetry group on the underlying spacetime manifold. For the class of metric-affine geometries, this notion of symmetry can be derived by realizing that they are particular classes of Cartan geometries, for which a more general notion of symmetry has been derived~\cite{Hohmann:2015pva}. This is the notion of symmetry we use in this article. It generalizes the more restricted notion for metric teleparallel geometries used in an earlier work~\cite{Hohmann:2019nat}. Here we focus on spherical symmetry, which is of particular importance, since it is typically assumed for non-rotating compact objects such as stars, black holes or exotic objects like wormholes.

The study of metric-affine geometries with spherical symmetry has a long history. For the most general metric-affine geometry featuring torsion, nonmetricity and curvature, explicit expressions have been obtained for a parametrization in terms of torsion and nonmetricity~\cite{Minkevich:2003it}. Particular spherically symmetric solutions of metric-affine gravity theories have been studied, e.g., in~\cite{Tresguerres:1995js,Tresguerres:1995un,Tucker:1995fw,Neeman:1996zcr}. For the case of Poincaré gauge theory, in which only torsion and curvature are present, spherically symmetric geometries have been studied, e.g., in~\cite{Bakler:1980is,Rauch:1981tva,Bakler:1983bm,Lee:1983af,Zhang:1985qc,Lenzen:1986hb,Bakler:1987uy,Ma:1994hq,Ma:1995cf,Mignemi:1996ge,Mignemi:1997hw,Babourova:2016gfd,Ziaie:2019dmq}. Solutions of Einstein-Cartan theory have been discussed, e.g., in~\cite{Som:1981zz,Singh:2002pk,Farfan:2011ic,Bronnikov:2015pha,Damour:2019oru}. For the case of teleparallel gravity, where only torsion is non-vanishing, see e.g.,~\cite{Ferraro:2011ks,Bohmer:2011si,Tamanini:2012hg,Paliathanasis:2014iva,Hohmann:2019nat,Boehmer:2019uxv,Bahamonde:2019zea}. Also for various other theories based on metric-affine geometry spherically symmetric solutions have been discussed~\cite{Damour:2002gp,Heinicke:2005bp,Eling:2006df,Filippov:2011aa,Cattani:2013dla,Babourova:2015bra,Babourova:2016pdh,Harris:2017nln,Olmo:2017qab}.

The aim of this article is twofold. Its primary aim is to provide a compendium of metric-affine geometries with spherical symmetry, ordered by the vanishing or non-vanishing of its characteristic tensorial properties - torsion, nonmetricity and curvature, and to provide explicit parametrizations which help to simplify the obtained expressions. These parametrized geometries may directly be inserted into the field equations of any gravity theory based on the corresponding subclass of metric-affine geometries, in order to find its spherically symmetric solutions. While such a parametrization has been provided already for the case of the symmetry group \(\mathrm{SO}(3)\) of proper rotations~\cite{Minkevich:2003it}, we extend this result to the full orthogonal group \(\mathrm{O}(3)\) including reflections. Another aim of this article is to demonstrate the method used for finding these spherically symmetric geometries, and thus to serve a didactic purpose. Using the same method it is possible to determine metric-affine geometries satisfying other spacetime symmetries, such as planar or cosmological symmetry. We briefly display also the most general metric-affine geometry with the latter kind of symmetry, to demonstrate how it follows from the spherically symmetric case which we study in detail.

We emphasize that it is \emph{not} the aim of this article to determine exact or approximate solutions to the field equations of any \emph{specific} gravity theory or class of gravity theories. We do \emph{not} choose any specific theory or class of theories in this article, or consider any field equations. The object of our study is given purely by the metric-affine geometries underlying such gravity theories, and by the action of spacetime symmetries - in particular the spherical symmetry - on these geometries. Choosing a gravity theory or class of theories and solving the corresponding field equations in spherical symmetry is an additional step, which is beyond the scope of this article.

The outline of this article is as follows. In section~\ref{sec:symmetry} we briefly review the notion of symmetry for metric-affine geometries, using both their metric-Palatini and tetrad / spin connection representations. We then derive the most general spherically symmetric metric-affine geometry in section~\ref{sec:spher}, where by spherical symmetry we consider symmetry under proper, orientation-preserving rotations only. The properties of this geometry are discussed in section~\ref{sec:props}. We consider particular subclasses in section~\ref{sec:special}, by imposing additional constraints on the connection. Our results are extended to the full orthogonal group including reflections in section~\ref{sec:reflect}. As a potential physical application, we study orbital motion along the autoparallels of the affine connection in section~\ref{sec:autoparallel}. Finally, we provide an outlook towards cosmological symmetry in section~\ref{sec:cosmo}, by displaying the most general metric-affine geometry obeying this symmetry. We end with a conclusion in section~\ref{sec:conclusion}.

\section{Symmetries of metric-affine geometries}\label{sec:symmetry}
The starting point of our derivation is the notion of spacetime symmetry for metric-affine geometries, which is derived from a more general notion of symmetry in Cartan geometry~\cite{Hohmann:2015pva}, and which we briefly review here. We do so in two formulations, based on different variables describing the geometry. We use a metric and an affine connection in section~\ref{ssec:metsym}, and a tetrad and a spin connection in section~\ref{ssec:tetsym}.

\subsection{Metric-Palatini formulation}\label{ssec:metsym}
We begin our review of symmetries of metric-affine geometries using the metric-Palatini formulation, according to which we will consider metric-affine geometries defined on a spacetime manifold \(M\) in terms of a metric \(g_{\mu\nu}\) and a connection with coefficients \(\Gamma^{\mu}{}_{\nu\rho}\). We use the convention in which the last index of the connection is the ``derivative index'', i.e., the covariant derivative of a vector field \(X^{\mu}\) in this convention reads \(\nabla_{\mu}X^{\nu} = \partial_{\mu}X^{\nu} + \Gamma^{\nu}{}_{\sigma\mu}X^{\sigma}\). This will be important later, since in general we will assume that the connection is not symmetric, i.e., it may possess torsion.

The notion of symmetry we use here is motivated by a previous study of symmetry in the language of Cartan geometry~\cite{Hohmann:2015pva} and its application to teleparallel gravity~\cite{Hohmann:2019nat}, which is a special case of the metric-affine geometries we consider here. In the following, we will consider the action \(\phi: G \times M \to M\) of a group \(G\) on the spacetime manifold \(M\). Denoting by \(\phi_u: M \to M\) for \(u \in G\) the induced diffeomorphism, which maps \(x \in M\) to \(\phi_u(x) = x'\), the metric and the connection coefficients transform as
\begin{equation}\label{eq:metsymcondf}
(\phi_u^*g)_{\mu\nu}(x) = g_{\rho\sigma}(x')\frac{\partial x'^{\rho}}{\partial x^{\mu}}\frac{\partial x'^{\sigma}}{\partial x^{\nu}}
\end{equation}
and
\begin{equation}\label{eq:affsymcondf}
(\phi_u^*\Gamma)^\mu{}_{\nu\rho}(x) = \Gamma^{\tau}{}_{\omega\sigma}(x')\frac{\partial x^{\mu}}{\partial x'^{\tau}}\frac{\partial x'^{\omega}}{\partial x^{\nu}}\frac{\partial x'^{\sigma}}{\partial x^{\rho}} + \frac{\partial x^{\mu}}{\partial x'^{\sigma}}\frac{\partial^2x'^{\sigma}}{\partial x^{\nu}\partial x^{\rho}}
\end{equation}
We call a metric-affine geometry \emph{symmetric} under this group action, if and only if for every \(u \in G\) the metric and affine connection are invariant, i.e., \((\phi_u^*g)_{\mu\nu} = g_{\mu\nu}\) and \((\phi_u^*\Gamma)^{\mu}{}_{\nu\rho} = \Gamma^{\mu}{}_{\nu\rho}\) for all \(u \in G\).

In practice, it is often more convenient to consider the infinitesimal action of the symmetry group, in terms of the generating (or fundamental) vector fields \(X_{\xi}\), where \(\xi \in \mathfrak{g}\) is an element of the Lie algebra of the symmetry group \(G\). The infinitesimal transformation of the metric-affine geometry is then defined by the Lie derivative
\begin{equation}\label{eq:metsymcondi}
(\mathcal{L}_{X_{\xi}}g)_{\mu\nu} = X_{\xi}^{\rho}\partial_{\rho}g_{\mu\nu} + \partial_{\mu}X_{\xi}^{\rho}g_{\rho\nu} + \partial_{\nu}X_{\xi}^{\rho}g_{\mu\rho}
\end{equation}
and
\begin{equation}\label{eq:affsymcondi}
\begin{split}
(\mathcal{L}_{X_{\xi}}\Gamma)^{\mu}{}_{\nu\rho} &= X_{\xi}^{\sigma}\partial_{\sigma}\Gamma^{\mu}{}_{\nu\rho} - \partial_{\sigma}X_{\xi}^{\mu}\Gamma^{\sigma}{}_{\nu\rho} + \partial_{\nu}X_{\xi}^{\sigma}\Gamma^{\mu}{}_{\sigma\rho} + \partial_{\rho}X_{\xi}^{\sigma}\Gamma^{\mu}{}_{\nu\sigma} + \partial_{\nu}\partial_{\rho}X_{\xi}^{\mu}\\
&= \nabla_{\rho}\nabla_{\nu}X_{\xi}^{\mu} - X_{\xi}^{\sigma}R^{\mu}{}_{\nu\rho\sigma} - \nabla_{\rho}(X_{\xi}^{\sigma}T^{\mu}{}_{\nu\sigma})\,,
\end{split}
\end{equation}
where the last line shows that the Lie derivative of the connection coefficients is actually a tensor, which can be expressed in terms of the curvature and torsion
\begin{equation}\label{eq:curvtors}
R^{\rho}{}_{\sigma\mu\nu} = \partial_{\mu}\Gamma^{\rho}{}_{\sigma\nu} - \partial_{\nu}\Gamma^{\rho}{}_{\sigma\mu} + \Gamma^{\rho}{}_{\tau\mu}\Gamma^{\tau}{}_{\sigma\nu} - \Gamma^{\rho}{}_{\tau\nu}\Gamma^{\tau}{}_{\sigma\mu}\,, \quad
T^{\rho}{}_{\mu\nu} = \Gamma^{\rho}{}_{\nu\mu} - \Gamma^{\rho}{}_{\mu\nu}
\end{equation}
of the connection, which we display here in order to clarify the conventions we will be using. We see that a metric-affine geometry is symmetric under a group action if the Lie derivative of the metric and the connection vanish for all generating vector fields, \((\mathcal{L}_{X_{\xi}}g)_{\mu\nu} = 0\) and \((\mathcal{L}_{X_{\xi}}\Gamma)^{\mu}{}_{\nu\rho} = 0\) for all \(\xi \in \mathfrak{g}\). This infinitesimal description is equivalent to the aforementioned description in terms of the group action, provided that the group is connected. For this reason, we will in the following consider spherical symmetry as invariance under the action of the pure rotation group \(\mathrm{SO}(3)\). A generalization to the full orthogonal group \(\mathrm{O}(3)\), which contains two connected components, and for which the purely infinitesimal treatment is insufficient, can be found in section~\ref{sec:reflect}.

\subsection{Tetrad / spin connection formulation}\label{ssec:tetsym}
Alternatively to the metric-Palatini formulation discussed above, one may also express the metric-affine geometry in terms of a tetrad \(\theta^a{}_{\mu}\) and a spin connection \(\omega^a{}_{b\mu}\), which define the metric and the affine connection coefficients through the relations
\begin{equation}\label{eq:metric}
g_{\mu\nu} = \eta_{ab}\theta^a{}_{\mu}\theta^b{}_{\nu}
\end{equation}
and
\begin{equation}\label{eq:affconn}
\Gamma^{\mu}{}_{\nu\rho} = e_a{}^{\mu}\left(\partial_{\rho}\theta^a{}_{\nu} + \omega^a{}_{b\rho}\theta^b{}_{\nu}\right)\,,
\end{equation}
where \(\eta_{ab} = \mathrm{diag}(-1,1,1,1)\) is the Minkowski metric and \(e_a{}^{\mu}\) is the inverse tetrad satisfying \(\theta^a{}_{\mu}e_b{}^{\mu} = \delta^a_b\) and \(\theta^a{}_{\mu}e_a{}^{\nu} = \delta_{\mu}^{\nu}\).

In order to understand the notion of symmetry in this formulation, first note that under a diffeomorphism the tetrad and the spin connection transform as one-forms,
\begin{equation}
(\phi_u^*\theta)^a{}_{\mu}(x) = \theta^a{}_{\nu}(x')\frac{\partial x'^{\nu}}{\partial x^{\mu}}\,, \quad
(\phi_u^*\omega)^a{}_{b\mu}(x) = \omega^a{}_{b\nu}(x')\frac{\partial x'^{\nu}}{\partial x^{\mu}}\,.
\end{equation}
Taking a look at the first relation, we find that the group action and the tetrad together define a map \(\boldsymbol{\Lambda}: G \times M \to \mathrm{GL}(4)\) such that
\begin{equation}
\boldsymbol{\Lambda}_u^a{}_b(x)(\phi_u^*\theta)^b{}_{\mu}(x) = \theta^a{}_{\mu}(x)
\end{equation}
for all \((u,x) \in G \times M\). In other words, \(\boldsymbol{\Lambda}_u(x) \in \mathrm{GL}(4)\) is the unique matrix which relates the transformed tetrad \((\phi_u^*\theta)^a{}_{\mu}\) and the original tetrad \(\theta^a{}_{\mu}\) at the point \(x\). It is now easy to see that \((\phi_u^*\theta)^a{}_{\mu}\) and \(\theta^a{}_{\mu}\) define the same metric if and only if they are related by a local Lorentz transformation, i.e., if and only if \(\boldsymbol{\Lambda}_u(x) \in \mathrm{SO}(1,3)\) for all \((u,x) \in G \times M\). Hence, we will consider a tetrad symmetric under the group action \(\phi\) if and only this condition is satisfied. Further, we find that \((\phi_u^*\Gamma)^{\mu}{}_{\nu\rho} = \Gamma^{\mu}{}_{\nu\rho}\) if and only if
\begin{equation}
(\phi_u^*\omega)^a{}_{b\mu} = (\boldsymbol{\Lambda}_u^{-1})^a{}_c\boldsymbol{\Lambda}_u^d{}_b\omega^c{}_{d\mu} + (\boldsymbol{\Lambda}_u^{-1})^a{}_c\partial_{\mu}\boldsymbol{\Lambda}_u^c{}_b\,.
\end{equation}
We can also express these conditions in an infinitesimal form. For this purpose we write the Lie derivative of the tetrad in the form
\begin{equation}\label{eq:tetlieder}
(\mathcal{L}_{X_{\xi}}\theta)^a{}_{\mu} = -\boldsymbol{\lambda}_{\xi}^a{}_b\theta^b{}_{\mu}\,,
\end{equation}
where \(\boldsymbol{\lambda}_{\xi}(x) \in \mathfrak{gl}(4)\) for a general (not necessarily symmetric) tetrad is given by
\begin{equation}\label{eq:lochomo}
\boldsymbol{\lambda}_{\xi}(x) = \left.\frac{\dd}{\dd t}\boldsymbol{\Lambda}_{\exp(t\xi)}(x)\right|_{t = 0}\,,
\end{equation}
and \(\exp: \mathfrak{g} \to G\) is the exponential map. In this case we find that the metric is symmetric, \((\mathcal{L}_{X_{\xi}}g)_{\mu\nu} = 0\), if and only if \(\boldsymbol{\lambda}_{\xi}(x) \in \mathfrak{so}(1,3)\) for all \((\xi,x) \in \mathfrak{g} \times M\). Further, we find that the Lie derivative~\eqref{eq:affsymcondi} of the connection~\eqref{eq:affconn} can be written as
\begin{equation}
(\mathcal{L}_{X_{\xi}}\Gamma)^{\mu}{}_{\nu\rho} = e_a{}^{\mu}\left[\theta^b{}_{\nu}(\mathcal{L}_{X_{\xi}}\omega)^a{}_{b\rho} + \DD_{\rho}(\mathcal{L}_{X_{\xi}}\theta)^a{}_{\nu}\right]
= e_a{}^{\mu}\theta^b{}_{\nu}\left[(\mathcal{L}_{X_{\xi}}\omega)^a{}_{b\rho} - \DD_{\rho}\boldsymbol{\lambda}_{\xi}^a{}_b\right]\,,
\end{equation}
where the second equality comes from the definition~\eqref{eq:tetlieder} and we introduced the total covariant derivative
\begin{subequations}
\begin{align}
\DD_{\mu}\boldsymbol{\lambda}_{\xi}^a{}_b &= \partial_{\mu}\boldsymbol{\lambda}_{\xi}^a{}_b + \omega^a{}_{c\mu}\boldsymbol{\lambda}_{\xi}^c{}_b - \omega^c{}_{b\mu}\boldsymbol{\lambda}_{\xi}^a{}_c\,,\\
\DD_{\mu}(\mathcal{L}_{X_{\xi}}\theta)^a{}_{\nu} &= \partial_{\mu}(\mathcal{L}_{X_{\xi}}\theta)^a{}_{\nu} + \omega^a{}_{b\mu}(\mathcal{L}_{X_{\xi}}\theta)^b{}_{\nu} - \Gamma^{\rho}{}_{\nu\mu}(\mathcal{L}_{X_{\xi}}\theta)^a{}_{\rho}\,,
\end{align}
\end{subequations}
which satisfies the ``tetrad postulate''
\begin{equation}
0 = \DD_{\mu}\theta^a{}_{\nu} = \partial_{\mu}\theta^a{}_{\nu} + \omega^a{}_{b\mu}\theta^b{}_{\nu} - \Gamma^{\rho}{}_{\nu\mu}\theta^a{}_{\rho}\,,
\end{equation}
as a consequence of the definition~\eqref{eq:affconn}. Hence, we find that the connection is symmetric under the action of the symmetry group if and only if the spin connection satisfies
\begin{equation}
(\mathcal{L}_{X_{\xi}}\omega)^a{}_{b\mu} = -e_b{}^{\nu}\DD_{\mu}(\mathcal{L}_{X_{\xi}}\theta)^a{}_{\nu} = \DD_{\rho}\boldsymbol{\lambda}_{\xi}^a{}_b
\end{equation}
for all \((\xi,x) \in \mathfrak{g} \times M\). This is the notion of symmetry we will apply in the following sections.

\section{Metric-affine geometry with spherical symmetry}\label{sec:spher}
After discussing the general notion of symmetry for a metric-affine geometry we now come to the particular case of spherical symmetry. We proceed in the following steps. We introduce the coordinates and conventions used for the symmetry generating vector fields in section~\ref{ssec:symgen}. We then briefly review the well-known derivation of the most general spherically symmetric metric in section~\ref{ssec:metric}, where we also introduce the convenient parametrization we will be using. For the connection, we derive the most general spherically symmetric coefficients in section~\ref{ssec:conn}. Finally, we discuss the tetrad and spin connection in section~\ref{ssec:tetspin}. Note that here we consider spherical symmetry as invariance under the action of the pure rotation group \(\mathrm{SO}(3)\); see section~\ref{sec:reflect} for a generalization to the full orthogonal group \(\mathrm{O}(3)\).

\subsection{Symmetry generating vector fields}\label{ssec:symgen}
In the following we will consider the symmetry of a metric-affine geometry under the rotation group, whose action on a spacetime with coordinates \((t,r,\vartheta,\varphi)\) is described by the generating vector fields
\begin{equation}\label{eq:genvecspher}
X_x = \sin\varphi\partial_{\vartheta} + \frac{\cos\varphi}{\tan\vartheta}\partial_{\varphi}\,, \quad
X_y = -\cos\varphi\partial_{\vartheta} + \frac{\sin\varphi}{\tan\vartheta}\partial_{\varphi}\,, \quad
X_z = -\partial_{\varphi}\,.
\end{equation}
They satisfy the commutation relations
\begin{equation}
[X_x, X_y] = X_z\,, \quad
[X_y, X_z] = X_x\,, \quad
[X_z, X_x] = X_y\,,
\end{equation}
which is the Lie algebra \(\mathfrak{so}(3)\) of the rotation group.

\subsection{Metric}\label{ssec:metric}
The most general metric with spherical symmetry is well known~\cite{Stephani:2003tm}. Here we briefly review its derivation, to exemplify the calculation which we will apply to the more involved case of the connection in the following section. First, demanding axial symmetry yields the condition
\begin{equation}
(\mathcal{L}_{X_z}g)_{\mu\nu} = -\partial_{\varphi}g_{\mu\nu} = 0\,,
\end{equation}
so that the metric components must be functions of the coordinates \(t,r,\vartheta\) only. In the next step, one considers the linear combination
\begin{equation}
\cos\varphi(\mathcal{L}_{X_x}g)_{\mu\nu} + \sin\varphi(\mathcal{L}_{X_y}g)_{\mu\nu} = 0\,.
\end{equation}
The advantage lies in the fact that the resulting equations are purely algebraic and take the form
\begin{equation}\label{eq:metricalgcond}
g_{t\vartheta} = g_{t\varphi} = g_{r\vartheta} = g_{r\varphi} = g_{\vartheta\varphi} = g_{\varphi\varphi} - g_{\vartheta\vartheta}\sin^2\vartheta = 0\,,
\end{equation}
and are easily solved. The remaining equations
\begin{equation}
\sin\varphi(\mathcal{L}_{X_x}g)_{\mu\nu} - \cos\varphi(\mathcal{L}_{X_y}g)_{\mu\nu} = 0
\end{equation}
then become similarly simple, and imply that the remaining independent components \(g_{tt}, g_{rr}, g_{tr}, g_{\vartheta\vartheta}\) must be functions of \(t\) and \(r\) only. Note that by a coordinate transformation one can still eliminate the component \(g_{tr}\)~\cite{Stephani:2003tm}; however, we will keep the general form here for later use. This will leave us the freedom to choose different coordinates, in which the metric exhibits off-diagonal components, but in which other fields under investigation may take a simpler form.

Also for later use we introduce the parametrization
\begin{equation}\label{eq:sphermetric}
g_{tt} = -e^{\mathcal{G}_1 + \mathcal{G}_2}\cos\mathcal{G}_3\,, \quad
g_{rr} = e^{\mathcal{G}_1 - \mathcal{G}_2}\cos\mathcal{G}_3\,, \quad
g_{tr} = e^{\mathcal{G}_1}\sin\mathcal{G}_3\,, \quad
g_{\vartheta\vartheta} = e^{\mathcal{G}_4}
\end{equation}
in terms of free functions \(\mathcal{G}_1(t,r), \ldots, \mathcal{G}_4(t,r)\), which will turn out to simplify various expressions we derive.

\subsection{Connection}\label{ssec:conn}
We then come to the connection. Demanding that the Lie derivative~\eqref{eq:affsymcondi} of the connection coefficients with respect to the vector field \(X_z\) vanishes, which simply reads \(-\partial_{\varphi}\Gamma^{\mu}{}_{\nu\rho}\), leads to the condition that they must be independent of \(\varphi\), hence they are functions of \(t,r,\vartheta\) only. To proceed with the remaining symmetry generating vector fields, as in the metric case it is helpful to consider first the linear combination
\begin{equation}
\cos\varphi(\mathcal{L}_{X_x}\Gamma)^{\mu}{}_{\nu\rho} + \sin\varphi(\mathcal{L}_{X_y}\Gamma)^{\mu}{}_{\nu\rho} = 0\,,
\end{equation}
which turns out to yield a set of purely algebraic equations. These equations take the form
\begin{gather}
\Gamma^t{}_{t\varphi} = \Gamma^t{}_{t\vartheta} = \Gamma^t{}_{r\varphi} = \Gamma^t{}_{r\vartheta} = \Gamma^r{}_{t\varphi} = \Gamma^r{}_{t\vartheta} = \Gamma^r{}_{r\varphi} = \Gamma^r{}_{r\vartheta} = 0\,,\nonumber\\
\Gamma^t{}_{\varphi t} = \Gamma^t{}_{\vartheta t} = \Gamma^t{}_{\varphi r} = \Gamma^t{}_{\vartheta r} = \Gamma^r{}_{\varphi t} = \Gamma^r{}_{\vartheta t} = \Gamma^r{}_{\varphi r} = \Gamma^r{}_{\vartheta r} = 0\,,\nonumber\\
\Gamma^{\varphi}{}_{tt} = \Gamma^{\vartheta}{}_{tt} = \Gamma^{\varphi}{}_{tr} = \Gamma^{\vartheta}{}_{tr} = \Gamma^{\varphi}{}_{rt} = \Gamma^{\vartheta}{}_{rt} = \Gamma^{\varphi}{}_{rr} = \Gamma^{\vartheta}{}_{rr} = 0\,,\\
\Gamma^{\vartheta}{}_{t\vartheta} - \Gamma^{\varphi}{}_{t\varphi} = \Gamma^{\vartheta}{}_{r\vartheta} - \Gamma^{\varphi}{}_{r\varphi} = \Gamma^{\vartheta}{}_{\vartheta t} - \Gamma^{\varphi}{}_{\varphi t} = \Gamma^{\vartheta}{}_{\vartheta r} - \Gamma^{\varphi}{}_{\varphi r} = \Gamma^t{}_{\vartheta\vartheta} - \frac{\Gamma^t{}_{\varphi\varphi}}{\sin^2\vartheta} = \Gamma^r{}_{\vartheta\vartheta} - \frac{\Gamma^r{}_{\varphi\varphi}}{\sin^2\vartheta} = 0\,,\nonumber\\
\Gamma^t{}_{\vartheta\varphi} + \Gamma^t{}_{\varphi\vartheta} = \Gamma^r{}_{\vartheta\varphi} + \Gamma^r{}_{\varphi\vartheta} = \Gamma^{\varphi}{}_{t\vartheta} + \frac{\Gamma^{\vartheta}{}_{t\varphi}}{\sin^2\vartheta} = \Gamma^{\varphi}{}_{r\vartheta} + \frac{\Gamma^{\vartheta}{}_{r\varphi}}{\sin^2\vartheta} = \Gamma^{\varphi}{}_{\vartheta t} + \frac{\Gamma^{\vartheta}{}_{\varphi t}}{\sin^2\vartheta} = \Gamma^{\varphi}{}_{\vartheta r} + \frac{\Gamma^{\vartheta}{}_{\varphi r}}{\sin^2\vartheta} = 0\,,\nonumber\\
\Gamma^{\vartheta}{}_{\vartheta\vartheta} = \Gamma^{\vartheta}{}_{\vartheta\varphi} = \Gamma^{\vartheta}{}_{\varphi\vartheta} = \Gamma^{\varphi}{}_{\vartheta\vartheta} = \Gamma^{\varphi}{}_{\varphi\varphi} = 0\,, \quad \Gamma^{\varphi}{}_{\vartheta\varphi} = \Gamma^{\varphi}{}_{\varphi\vartheta} = \cot\vartheta\,, \quad \Gamma^{\vartheta}{}_{\varphi\varphi} = -\sin\vartheta\cos\vartheta\,,\nonumber
\end{gather}
and determine 44 of the 64 components of the connection coefficients in terms of the remaining 20 components. These components are further constrained by imposing the remaining linear combination
\begin{equation}
\sin\varphi(\mathcal{L}_{X_x}\Gamma)^{\mu}{}_{\nu\rho} - \cos\varphi(\mathcal{L}_{X_y}\Gamma)^{\mu}{}_{\nu\rho} = 0
\end{equation}
of the symmetry conditions, which take the form
\begin{gather}
\partial_{\vartheta}\Gamma^t{}_{tt} = \partial_{\vartheta}\Gamma^t{}_{tr} = \partial_{\vartheta}\Gamma^t{}_{rt} = \partial_{\vartheta}\Gamma^t{}_{rr} = \partial_{\vartheta}\Gamma^r{}_{tt} = \partial_{\vartheta}\Gamma^r{}_{tr} = \partial_{\vartheta}\Gamma^r{}_{rt} = \partial_{\vartheta}\Gamma^r{}_{rr} = 0\,,\nonumber\\
\partial_{\vartheta}\Gamma^{\varphi}{}_{t\varphi} = \partial_{\vartheta}\Gamma^{\varphi}{}_{r\varphi} = \partial_{\vartheta}\Gamma^{\varphi}{}_{\varphi t} = \partial_{\vartheta}\Gamma^{\varphi}{}_{\varphi r} = 0\,,\\
\partial_{\vartheta}\Gamma^{\varphi}{}_{t\vartheta} + \cot\vartheta\Gamma^{\varphi}{}_{t\vartheta} = \partial_{\vartheta}\Gamma^{\varphi}{}_{r\vartheta} + \cot\vartheta\Gamma^{\varphi}{}_{r\vartheta} = \partial_{\vartheta}\Gamma^{\varphi}{}_{\vartheta t} + \cot\vartheta\Gamma^{\varphi}{}_{\vartheta t} = \partial_{\vartheta}\Gamma^{\varphi}{}_{\vartheta r} + \cot\vartheta\Gamma^{\varphi}{}_{\vartheta r} = 0\,,\nonumber\\
\partial_{\vartheta}\Gamma^t{}_{\varphi\varphi} - 2\cot\vartheta\Gamma^t{}_{\varphi\varphi} = \partial_{\vartheta}\Gamma^r{}_{\varphi\varphi} - 2\cot\vartheta\Gamma^r{}_{\varphi\varphi} = \partial_{\vartheta}\Gamma^t{}_{\varphi\vartheta} - \cot\vartheta\Gamma^t{}_{\varphi\vartheta} = \partial_{\vartheta}\Gamma^r{}_{\varphi\vartheta} - \cot\vartheta\Gamma^r{}_{\varphi\vartheta} = 0\,.\nonumber
\end{gather}
We find a number of differential equations, which fully determine the dependence of the remaining components on the coordinate \(\vartheta\). They can be solved explicitly, and their solution is expressed in terms of 20 functions \(\mathcal{C}_1(t,r), \ldots, \mathcal{C}_{20}(t,r)\) of the remaining coordinates \(t,r\) as
\begin{gather}
\Gamma^t{}_{tt} = \mathcal{C}_1\,, \quad
\Gamma^t{}_{tr} = \mathcal{C}_2\,, \quad
\Gamma^t{}_{rt} = \mathcal{C}_3\,, \quad
\Gamma^t{}_{rr} = \mathcal{C}_4\,, \quad
\Gamma^t{}_{\vartheta\vartheta} = \mathcal{C}_9\,,\nonumber\\
\Gamma^r{}_{tt} = \mathcal{C}_5\,, \quad
\Gamma^r{}_{tr} = \mathcal{C}_6\,, \quad
\Gamma^r{}_{rt} = \mathcal{C}_7\,, \quad
\Gamma^r{}_{rr} = \mathcal{C}_8\,, \quad
\Gamma^r{}_{\vartheta\vartheta} = \mathcal{C}_{10}\,,\nonumber\\
\Gamma^{\varphi}{}_{t\varphi} = \Gamma^{\vartheta}{}_{t\vartheta} = \mathcal{C}_{11}\,, \quad
\Gamma^{\varphi}{}_{r\varphi} = \Gamma^{\vartheta}{}_{r\vartheta} = \mathcal{C}_{12}\,, \quad
\Gamma^{\varphi}{}_{\varphi t} = \Gamma^{\vartheta}{}_{\vartheta t} = \mathcal{C}_{13}\,, \quad
\Gamma^{\varphi}{}_{\varphi r} = \Gamma^{\vartheta}{}_{\vartheta r} = \mathcal{C}_{14}\,,\nonumber\\
\Gamma^{\varphi}{}_{t\vartheta} = \frac{\mathcal{C}_{15}}{\sin\vartheta}\,, \quad
\Gamma^{\vartheta}{}_{t\varphi} = -\mathcal{C}_{15}\sin\vartheta\,, \quad
\Gamma^{\varphi}{}_{r\vartheta} = \frac{\mathcal{C}_{16}}{\sin\vartheta}\,, \quad
\Gamma^{\vartheta}{}_{r\varphi} = -\mathcal{C}_{16}\sin\vartheta\,,\label{eq:sphergamma}\\
\Gamma^{\varphi}{}_{\vartheta t} = \frac{\mathcal{C}_{17}}{\sin\vartheta}\,, \quad
\Gamma^{\vartheta}{}_{\varphi t} = -\mathcal{C}_{17}\sin\vartheta\,, \quad
\Gamma^{\varphi}{}_{\vartheta r} = \frac{\mathcal{C}_{18}}{\sin\vartheta}\,, \quad
\Gamma^{\vartheta}{}_{\varphi r} = -\mathcal{C}_{18}\sin\vartheta\,,\nonumber\\
\Gamma^t{}_{\varphi\vartheta} = \mathcal{C}_{19}\sin\vartheta\,, \quad
\Gamma^t{}_{\vartheta\varphi} = -\mathcal{C}_{19}\sin\vartheta\,, \quad
\Gamma^r{}_{\varphi\vartheta} = \mathcal{C}_{20}\sin\vartheta\,, \quad
\Gamma^r{}_{\vartheta\varphi} = -\mathcal{C}_{20}\sin\vartheta\,,\nonumber\\
\Gamma^t{}_{\varphi\varphi} = \mathcal{C}_9\sin^2\vartheta\,, \quad
\Gamma^r{}_{\varphi\varphi} = \mathcal{C}_{10}\sin^2\vartheta\,, \quad
\Gamma^{\varphi}{}_{\vartheta\varphi} = \Gamma^{\varphi}{}_{\varphi\vartheta} = \cot\vartheta\,, \quad
\Gamma^{\vartheta}{}_{\varphi\varphi} = -\sin\vartheta\cos\vartheta\,.\nonumber
\end{gather}
One easily checks that these satisfy the symmetry conditions.

\subsection{Tetrad and spin connection}\label{ssec:tetspin}
In order to construct a tetrad and spin connection which satisfy the conditions of spherical symmetry, one may start from the metric~\eqref{eq:sphermetric} and affine connection~\eqref{eq:sphergamma} derived in the previous section. For the tetrad it is sufficient to choose any tetrad for which the metric~\eqref{eq:metric} obeys the spherical symmetry. One may thus choose, e.g., the tetrad
\begin{align}
\theta^0 &= e^{\tilde{\mathcal{G}}_1 + \tilde{\mathcal{G}}_2}\cos\tilde{\mathcal{G}}_3\,\dd t - e^{\tilde{\mathcal{G}}_1 - \tilde{\mathcal{G}}_2}\sin\tilde{\mathcal{G}}_3\,\dd r\,, &
\theta^2 &= e^{\tilde{\mathcal{G}}_4}\,\dd\vartheta\,,\nonumber\\
\theta^1 &= e^{\tilde{\mathcal{G}}_1 + \tilde{\mathcal{G}}_2}\sin\tilde{\mathcal{G}}_3\,\dd t + e^{\tilde{\mathcal{G}}_1 - \tilde{\mathcal{G}}_2}\cos\tilde{\mathcal{G}}_3\,\dd r\,, &
\theta^3 &= e^{\tilde{\mathcal{G}}_4}\sin\vartheta\,\dd\varphi\,,
\end{align}
where we wrote \(\tilde{\mathcal{G}}_i = \mathcal{G}_i/2\) for brevity. Inserting this tetrad in the definition~\eqref{eq:affconn} and solving for the spin connection then yields the non-vanishing components
\begin{gather}
\omega^0{}_{0t} = \mathcal{S}_1\,, \quad
\omega^0{}_{0r} = \mathcal{S}_2\,, \quad
\omega^0{}_{1t} = \mathcal{S}_3\,, \quad
\omega^0{}_{1r} = \mathcal{S}_4\,, \quad
\omega^3{}_{2\varphi} = \cos\vartheta\,,\nonumber\\
\omega^1{}_{0t} = \mathcal{S}_5\,, \quad
\omega^1{}_{0r} = \mathcal{S}_6\,, \quad
\omega^1{}_{1t} = \mathcal{S}_7\,, \quad
\omega^1{}_{1r} = \mathcal{S}_8\,, \quad
\omega^2{}_{3\varphi} = -\cos\vartheta\,,\nonumber\\
\omega^0{}_{2\vartheta} = \frac{\omega^0{}_{3\varphi}}{\sin\vartheta} = \mathcal{S}_9\,, \quad
\omega^1{}_{2\vartheta} = \frac{\omega^1{}_{3\varphi}}{\sin\vartheta} = \mathcal{S}_{10}\,, \quad
\omega^2{}_{0\vartheta} = \frac{\omega^3{}_{0\varphi}}{\sin\vartheta} = \mathcal{S}_{11}\,, \quad
\omega^2{}_{1\vartheta} = \frac{\omega^3{}_{1\varphi}}{\sin\vartheta} = \mathcal{S}_{12}\,,\\
\omega^2{}_{2t} = \omega^3{}_{3t} = \mathcal{S}_{13}\,, \quad
\omega^2{}_{2r} = \omega^3{}_{3r} = \mathcal{S}_{14}\,, \quad
\omega^3{}_{0\vartheta} = -\frac{\omega^2{}_{0\varphi}}{\sin\vartheta} = \mathcal{S}_{15}\,, \quad
\omega^3{}_{1\vartheta} = -\frac{\omega^2{}_{1\varphi}}{\sin\vartheta} = \mathcal{S}_{16}\,,\nonumber\\
\omega^3{}_{2t} = -\omega^2{}_{3t} = \mathcal{S}_{17}\,, \quad
\omega^3{}_{2r} = -\omega^2{}_{3r} = \mathcal{S}_{18}\,, \quad
\omega^0{}_{3\vartheta} = -\frac{\omega^0{}_{2\varphi}}{\sin\vartheta} = \mathcal{S}_{19}\,, \quad
\omega^1{}_{3\vartheta} = -\frac{\omega^1{}_{2\varphi}}{\sin\vartheta} = \mathcal{S}_{20}\,,\nonumber
\end{gather}
where the different parametrizations are related by
\begin{gather}
\mathcal{C}_1 = \mathcal{S}_1\cos^2\tilde{\mathcal{G}}_3 + \frac{1}{2}(\mathcal{S}_3 + \mathcal{S}_5)\sin(2\tilde{\mathcal{G}}_3) + \mathcal{S}_7\sin^2\tilde{\mathcal{G}}_3 + \tilde{\mathcal{G}}_{1,t} + \tilde{\mathcal{G}}_{2,t}\,,\nonumber\\
\mathcal{C}_2 = \mathcal{S}_2\cos^2\tilde{\mathcal{G}}_3 + \frac{1}{2}(\mathcal{S}_4 + \mathcal{S}_6)\sin(2\tilde{\mathcal{G}}_3) + \mathcal{S}_8\sin^2\tilde{\mathcal{G}}_3 + \tilde{\mathcal{G}}_{1,r} + \tilde{\mathcal{G}}_{2,r}\,,\nonumber\\
\mathcal{C}_3 = \left[\mathcal{S}_3\cos^2\tilde{\mathcal{G}}_3 + \frac{1}{2}(\mathcal{S}_7 - \mathcal{S}_1)\sin(2\tilde{\mathcal{G}}_3) - \mathcal{S}_5\sin^2\tilde{\mathcal{G}}_3 - \tilde{\mathcal{G}}_{3,t}\right]e^{-2\tilde{\mathcal{G}}_2}\,,\nonumber\\
\mathcal{C}_4 = \left[\mathcal{S}_4\cos^2\tilde{\mathcal{G}}_3 + \frac{1}{2}(\mathcal{S}_8 - \mathcal{S}_2)\sin(2\tilde{\mathcal{G}}_3) - \mathcal{S}_6\sin^2\tilde{\mathcal{G}}_3 - \tilde{\mathcal{G}}_{3,r}\right]e^{-2\tilde{\mathcal{G}}_2}\,,\nonumber\\
\mathcal{C}_5 = \left[\mathcal{S}_5\cos^2\tilde{\mathcal{G}}_3 + \frac{1}{2}(\mathcal{S}_7 - \mathcal{S}_1)\sin(2\tilde{\mathcal{G}}_3) - \mathcal{S}_3\sin^2\tilde{\mathcal{G}}_3 + \tilde{\mathcal{G}}_{3,t}\right]e^{2\tilde{\mathcal{G}}_2}\,,\nonumber\\
\mathcal{C}_6 = \left[\mathcal{S}_6\cos^2\tilde{\mathcal{G}}_3 + \frac{1}{2}(\mathcal{S}_8 - \mathcal{S}_2)\sin(2\tilde{\mathcal{G}}_3) - \mathcal{S}_4\sin^2\tilde{\mathcal{G}}_3 + \tilde{\mathcal{G}}_{3,r}\right]e^{2\tilde{\mathcal{G}}_2}\,,\nonumber\\
\mathcal{C}_7 = \mathcal{S}_7\cos^2\tilde{\mathcal{G}}_3 - \frac{1}{2}(\mathcal{S}_3 + \mathcal{S}_5)\sin(2\tilde{\mathcal{G}}_3) + \mathcal{S}_1\sin^2\tilde{\mathcal{G}}_3 + \tilde{\mathcal{G}}_{1,t} - \tilde{\mathcal{G}}_{2,t}\,,\nonumber\\
\mathcal{C}_8 = \mathcal{S}_8\cos^2\tilde{\mathcal{G}}_3 - \frac{1}{2}(\mathcal{S}_4 + \mathcal{S}_6)\sin(2\tilde{\mathcal{G}}_3) + \mathcal{S}_2\sin^2\tilde{\mathcal{G}}_3 + \tilde{\mathcal{G}}_{1,r} - \tilde{\mathcal{G}}_{2,r}\,,\nonumber\\
\mathcal{C}_9 = (\mathcal{S}_{10}\sin\tilde{\mathcal{G}}_3 + \mathcal{S}_9\cos\tilde{\mathcal{G}}_3)e^{\tilde{\mathcal{G}}_4 - \tilde{\mathcal{G}}_1 - \tilde{\mathcal{G}}_2}\,, \quad
\mathcal{C}_{11} = (\mathcal{S}_{11}\cos\tilde{\mathcal{G}}_3 - \mathcal{S}_{12}\sin\tilde{\mathcal{G}}_3)e^{\tilde{\mathcal{G}}_1 + \tilde{\mathcal{G}}_2 - \tilde{\mathcal{G}}_4}\,,\nonumber\\
\mathcal{C}_{10} = (\mathcal{S}_{10}\cos\tilde{\mathcal{G}}_3 - \mathcal{S}_9\sin\tilde{\mathcal{G}}_3)e^{\tilde{\mathcal{G}}_4 - \tilde{\mathcal{G}}_1 + \tilde{\mathcal{G}}_2}\,, \quad
\mathcal{C}_{12} = (\mathcal{S}_{11}\sin\tilde{\mathcal{G}}_3 + \mathcal{S}_{12}\cos\tilde{\mathcal{G}}_3)e^{\tilde{\mathcal{G}}_1 - \tilde{\mathcal{G}}_2 - \tilde{\mathcal{G}}_4}\,,\nonumber\\
\mathcal{C}_{19} = (\mathcal{S}_{20}\sin\tilde{\mathcal{G}}_3 + \mathcal{S}_{19}\cos\tilde{\mathcal{G}}_3)e^{\tilde{\mathcal{G}}_4 - \tilde{\mathcal{G}}_1 - \tilde{\mathcal{G}}_2}\,, \quad
\mathcal{C}_{15} = (\mathcal{S}_{15}\cos\tilde{\mathcal{G}}_3 - \mathcal{S}_{16}\sin\tilde{\mathcal{G}}_3)e^{\tilde{\mathcal{G}}_1 + \tilde{\mathcal{G}}_2 - \tilde{\mathcal{G}}_4}\,,\nonumber\\
\mathcal{C}_{20} = (\mathcal{S}_{20}\cos\tilde{\mathcal{G}}_3 - \mathcal{S}_{19}\sin\tilde{\mathcal{G}}_3)e^{\tilde{\mathcal{G}}_4 - \tilde{\mathcal{G}}_1 + \tilde{\mathcal{G}}_2}\,, \quad
\mathcal{C}_{16} = (\mathcal{S}_{15}\sin\tilde{\mathcal{G}}_3 + \mathcal{S}_{16}\cos\tilde{\mathcal{G}}_3)e^{\tilde{\mathcal{G}}_1 - \tilde{\mathcal{G}}_2 - \tilde{\mathcal{G}}_4}\,,\nonumber\\
\mathcal{C}_{13} = \mathcal{S}_{13} + \tilde{\mathcal{G}}_{4,t}\,, \quad
\mathcal{C}_{14} = \mathcal{S}_{14} + \tilde{\mathcal{G}}_{4,r}\,, \quad
\mathcal{C}_{17} = \mathcal{S}_{17}\,, \quad
\mathcal{C}_{18} = \mathcal{S}_{18}\,.
\end{gather}
This generalizes the spin connection found in~\cite{Minkevich:2003it} for a diagonal tetrad. Of course, every other tetrad and spin connection which are related to the original tetrad and spin connection by a local Lorentz transformation \(\Lambda: M \to \mathrm{SO}(1,3)\) via
\begin{equation}
\theta'^a{}_{\mu} = \Lambda^a{}_b\theta^b{}_{\mu}\,, \quad
\omega'^a{}_{b\mu} = \Lambda^a{}_c(\Lambda^{-1})^d{}_b\omega^c{}_{d\mu} + \Lambda^a{}_c\partial_{\mu}(\Lambda^{-1})^c{}_b
\end{equation}
also satisfies the conditions of spherical symmetry. This allows to obtain different, alternative representations of the spherical metric-affine geometry. Here we restrict ourselves to listing only one example, since the tensorial quantities, which we will calculate in the following sections, are independent of this choice. See, e.g.,~\cite{Hohmann:2019nat} for a number of alternative tetrad representations of the spherically symmetric metric and the Lorentz transformations relating these different tetrads.

\section{Properties of the geometry}\label{sec:props}
We now discuss a number of properties of the most general spherically symmetric metric-affine geometry, which we derived in the preceding section. In particular, we calculate its torsion in section~\ref{ssec:torsion} and its nonmetricity in section~\ref{ssec:nonmet}. This will lead us to the decomposition of the connection into the Levi-Civita connection, the contortion and disformation in section~\ref{ssec:conndecomp}. Finally, we discuss the curvature in section~\ref{ssec:curv}

\subsection{Torsion}\label{ssec:torsion}
We start by discussing the torsion \(T^{\mu}{}_{\nu\rho} = \Gamma^{\mu}{}_{\rho\nu} - \Gamma^{\mu}{}_{\nu\rho}\) of the most general spherically symmetric connection~\eqref{eq:sphergamma}. We find that its non-vanishing, independent components are given by
\begin{align}
T^t{}_{tr} &= \mathcal{C}_3 - \mathcal{C}_2\,, &
T^t{}_{\vartheta\varphi} &= 2\mathcal{C}_{19}\sin\vartheta\,, &
T^{\vartheta}{}_{t\vartheta} &= T^{\varphi}{}_{t\varphi} = \mathcal{C}_{13} - \mathcal{C}_{11}\,, &
T^{\varphi}{}_{t\vartheta} &= -\frac{T^{\vartheta}{}_{t\varphi}}{\sin^2\vartheta} = \frac{\mathcal{C}_{17} - \mathcal{C}_{15}}{\sin\vartheta}\,,\nonumber\\
T^r{}_{tr} &= \mathcal{C}_7 - \mathcal{C}_6\,, &
T^r{}_{\vartheta\varphi} &= 2\mathcal{C}_{20}\sin\vartheta\,, &
T^{\vartheta}{}_{r\vartheta} &= T^{\varphi}{}_{r\varphi} = \mathcal{C}_{14} - \mathcal{C}_{12}\,, &
T^{\varphi}{}_{r\vartheta} &= -\frac{T^{\vartheta}{}_{r\varphi}}{\sin^2\vartheta} = \frac{\mathcal{C}_{18} - \mathcal{C}_{16}}{\sin\vartheta}\,.\label{eq:torsionc}
\end{align}
Note that the torsion depends only on 8 particular combinations of the parameter functions \(\mathcal{C}_1, \ldots, \mathcal{C}_{20}\). This will become relevant in section~\ref{ssec:conndecomp}, when we decompose the connection.

\subsection{Nonmetricity}\label{ssec:nonmet}
We then continue with the nonmetricity \(Q_{\mu\nu\rho} = \nabla_{\mu}g_{\nu\rho}\). The non-vanishing, independent components for the metric~\eqref{eq:sphermetric} and the connection~\eqref{eq:sphergamma} take the form
\begin{align}
Q_{ttt} &= -e^{\mathcal{G}_1}\left[\left(2\mathcal{C}_5 - e^{\mathcal{G}_2}\mathcal{G}_{3,t}\right)\sin\mathcal{G}_3 - \left(2\mathcal{C}_1 - \mathcal{G}_{1,t} - \mathcal{G}_{2,t}\right)e^{\mathcal{G}_2}\cos\mathcal{G}_3\right]\,,\nonumber\\
Q_{rtt} &= -e^{\mathcal{G}_1}\left[\left(2\mathcal{C}_6 - e^{\mathcal{G}_2}\mathcal{G}_{3,r}\right)\sin\mathcal{G}_3 - \left(2\mathcal{C}_2 - \mathcal{G}_{1,r} - \mathcal{G}_{2,r}\right)e^{\mathcal{G}_2}\cos\mathcal{G}_3\right]\,,\nonumber\\
Q_{trr} &= -e^{\mathcal{G}_1}\left[\left(2\mathcal{C}_3 + e^{-\mathcal{G}_2}\mathcal{G}_{3,t}\right)\sin\mathcal{G}_3 + \left(2\mathcal{C}_7 - \mathcal{G}_{1,t} + \mathcal{G}_{2,t}\right)e^{-\mathcal{G}_2}\cos\mathcal{G}_3\right]\,,\nonumber\\
Q_{rrr} &= -e^{\mathcal{G}_1}\left[\left(2\mathcal{C}_4 + e^{-\mathcal{G}_2}\mathcal{G}_{3,r}\right)\sin\mathcal{G}_3 + \left(2\mathcal{C}_8 - \mathcal{G}_{1,r} + \mathcal{G}_{2,r}\right)e^{-\mathcal{G}_2}\cos\mathcal{G}_3\right]\,,\nonumber\\
Q_{ttr} &= -e^{\mathcal{G}_1}\left[\left(\mathcal{C}_1 + \mathcal{C}_7 - \mathcal{G}_{1,t}\right)\sin\mathcal{G}_3 - \left(\mathcal{C}_3e^{\mathcal{G}_2} - \mathcal{C}_5e^{-\mathcal{G}_2} + \mathcal{G}_{3,t}\right)\cos\mathcal{G}_3\right]\,,\nonumber\\
Q_{rtr} &= -e^{\mathcal{G}_1}\left[\left(\mathcal{C}_2 + \mathcal{C}_8 - \mathcal{G}_{1,r}\right)\sin\mathcal{G}_3 - \left(\mathcal{C}_4e^{\mathcal{G}_2} - \mathcal{C}_6e^{-\mathcal{G}_2} + \mathcal{G}_{3,r}\right)\cos\mathcal{G}_3\right]\,,\nonumber\\
Q_{t\vartheta\vartheta} &= \frac{Q_{t\varphi\varphi}}{\sin^2\vartheta} = e^{\mathcal{G}_4}\left(\mathcal{G}_{4,t} - 2\mathcal{C}_{13}\right)\,,\nonumber\\
Q_{r\vartheta\vartheta} &= \frac{Q_{r\varphi\varphi}}{\sin^2\vartheta} = e^{\mathcal{G}_4}\left(\mathcal{G}_{4,r} - 2\mathcal{C}_{14}\right)\,,\nonumber\\
Q_{\varphi t\vartheta} &= -Q_{\vartheta t\varphi} = \left[\mathcal{C}_{15}e^{\mathcal{G}_4} + \left(\mathcal{C}_{20}\sin\mathcal{G}_3 - \mathcal{C}_{19}e^{\mathcal{G}_2}\cos\mathcal{G}_3\right)e^{\mathcal{G}_1}\right]\sin\vartheta\,,\nonumber\\
Q_{\varphi r\vartheta} &= -Q_{\vartheta r\varphi} = \left[\mathcal{C}_{16}e^{\mathcal{G}_4} + \left(\mathcal{C}_{19}\sin\mathcal{G}_3 + \mathcal{C}_{20}e^{-\mathcal{G}_2}\cos\mathcal{G}_3\right)e^{\mathcal{G}_1}\right]\sin\vartheta\,,\nonumber\\
Q_{\vartheta t\vartheta} &= \frac{Q_{\varphi t\varphi}}{\sin^2\vartheta} = -\mathcal{C}_{11}e^{\mathcal{G}_4} - \left(\mathcal{C}_{10}\sin\mathcal{G}_3 - \mathcal{C}_9e^{\mathcal{G}_2}\cos\mathcal{G}_3\right)e^{\mathcal{G}_1}\,,\nonumber\\
Q_{\vartheta r\vartheta} &= \frac{Q_{\varphi r\varphi}}{\sin^2\vartheta} = -\mathcal{C}_{12}e^{\mathcal{G}_4} - \left(\mathcal{C}_9\sin\mathcal{G}_3 + \mathcal{C}_{10}e^{-\mathcal{G}_2}\cos\mathcal{G}_3\right)e^{\mathcal{G}_1}\,.\label{eq:nonmetc}
\end{align}
In total, we have 12 independent components. The significance of this observation becomes clear in the following section.

\subsection{Connection decomposition}\label{ssec:conndecomp}
We now make use of the fact that the coefficients of an arbitrary connection can uniquely be decomposed in the form
\begin{equation}\label{eq:conndecomp}
\Gamma^{\mu}{}_{\nu\rho} = \lc{\Gamma}^{\mu}{}_{\nu\rho} + K^{\mu}{}_{\nu\rho} + L^{\mu}{}_{\nu\rho}\,,
\end{equation}
where \(\lc{\Gamma}^{\mu}{}_{\nu\rho}\) is the Levi-Civita connection of the metric \(g_{\mu\nu}\), \(K^{\mu}{}_{\nu\rho}\) is the contortion
\begin{equation}\label{eq:contortion}
K^{\mu}{}_{\nu\rho} = \frac{1}{2}\left(T_{\nu}{}^{\mu}{}_{\rho} + T_{\rho}{}^{\mu}{}_{\nu} - T^{\mu}{}_{\nu\rho}\right)
\end{equation}
and \(L^{\mu}{}_{\nu\rho}\) is the disformation
\begin{equation}\label{eq:disformation}
L^{\mu}{}_{\nu\rho} = \frac{1}{2}\left(Q^{\mu}{}_{\nu\rho} - Q_{\nu}{}^{\mu}{}_{\rho} - Q_{\rho}{}^{\mu}{}_{\nu}\right)\,.
\end{equation}
In order to decompose the connection~\eqref{eq:sphergamma} with respect to the metric~\eqref{eq:sphermetric}, it is useful to introduce a different parametrization in terms of free functions \(\mathcal{T}_1(t, r), \ldots, \mathcal{T}_8(t, r)\) and \(\mathcal{Q}_1(t, r), \ldots, \mathcal{Q}_{12}(t, r)\); see~\cite{Minkevich:2003it} for a similar parametrization. We replace the previously introduced parameters by making the substitutions
\begin{gather}
\mathcal{C}_1 = \frac{1}{2}e^{-\mathcal{G}_1}\left[\mathcal{Q}_2e^{-\mathcal{G}_2}\cos\mathcal{G}_3 + (\mathcal{Q}_5 - 2\mathcal{Q}_3)\sin\mathcal{G}_3\right] + \frac{1}{4}\left[\left(2\mathcal{T}_1 + \mathcal{G}_{1,r} + \mathcal{G}_{2,r}\right)e^{\mathcal{G}_2} + \mathcal{G}_{3,t}\right]\sin(2\mathcal{G}_3)\nonumber\\
+ \frac{1}{4}\mathcal{G}_{1,t}(3 - \cos(2\mathcal{G}_3)) + \frac{1}{2}\mathcal{G}_{2,t}\cos^2\mathcal{G}_3 - \frac{1}{2}\left(2\mathcal{T}_2 + \mathcal{G}_{3,r}e^{\mathcal{G}_2}\right)\sin^2\mathcal{G}_3\,,\nonumber\\
\mathcal{C}_3 = \frac{1}{2}e^{-\mathcal{G}_1}\left(\mathcal{Q}_5e^{-\mathcal{G}_2}\cos\mathcal{G}_3 - \mathcal{Q}_2\sin\mathcal{G}_3\right) - \frac{1}{4}\left[\left(2\mathcal{T}_2 - \mathcal{G}_{1,t} + \mathcal{G}_{2,t}\right)e^{-\mathcal{G}_2} + \mathcal{G}_{3,r}\right]\sin(2\mathcal{G}_3)\nonumber\\
- \frac{1}{2}\mathcal{G}_{3,t}e^{-\mathcal{G}_2}\sin^2\mathcal{G}_3 + \frac{1}{2}\left(\mathcal{G}_{1,r} + \mathcal{G}_{2,r} + 2\mathcal{T}_1\right)\cos^2\mathcal{G}_3\,,\nonumber\\
\mathcal{C}_4 = \frac{1}{2}e^{-\mathcal{G}_1}\left[(2\mathcal{Q}_7 - \mathcal{Q}_2)e^{-\mathcal{G}_2}\cos\mathcal{G}_3 - \mathcal{Q}_6\sin\mathcal{G}_3\right] - \frac{1}{4}\left(2\mathcal{T}_1 + \mathcal{G}_{1,r} + \mathcal{G}_{2,r} + \mathcal{G}_{3,t}e^{-\mathcal{G}_2}\right)e^{-\mathcal{G}_2}\sin(2\mathcal{G}_3)\nonumber\\
- \frac{1}{4}\mathcal{G}_{3,r}e^{-\mathcal{G}_2}(3 + \cos(2\mathcal{G}_3)) + \frac{1}{2}\left(\mathcal{G}_{1,t} - \mathcal{G}_{2,t} - 2\mathcal{T}_2\right)e^{-2\mathcal{G}_2}\cos^2\mathcal{G}_3\,,\nonumber\\
\mathcal{C}_5 = \frac{1}{2}e^{-\mathcal{G}_1}\left[(\mathcal{Q}_5 - 2\mathcal{Q}_3)e^{\mathcal{G}_2}\cos\mathcal{G}_3 - \mathcal{Q}_1\sin\mathcal{G}_3\right] - \frac{1}{4}\left(2\mathcal{T}_2 - \mathcal{G}_{1,t} + \mathcal{G}_{2,t} + \mathcal{G}_{3,r}e^{\mathcal{G}_2}\right)e^{\mathcal{G}_2}\sin(2\mathcal{G}_3)\nonumber\\
+ \frac{1}{4}\mathcal{G}_{3,t}e^{\mathcal{G}_2}(3 + \cos(2\mathcal{G}_3)) + \frac{1}{2}\left(\mathcal{G}_{1,r} + \mathcal{G}_{2,r} + 2\mathcal{T}_1\right)e^{2\mathcal{G}_2}\cos^2\mathcal{G}_3\,,\nonumber\\
\mathcal{C}_6 = -\frac{1}{2}e^{-\mathcal{G}_1}\left(\mathcal{Q}_2e^{\mathcal{G}_2}\cos\mathcal{G}_3 + \mathcal{Q}_5\sin\mathcal{G}_3\right) - \frac{1}{4}\left[\left(2\mathcal{T}_1 + \mathcal{G}_{1,r} + \mathcal{G}_{2,r}\right)e^{\mathcal{G}_2} + \mathcal{G}_{3,t}\right]\sin(2\mathcal{G}_3)\nonumber\\
+ \frac{1}{2}\mathcal{G}_{3,r}e^{\mathcal{G}_2}\sin^2\mathcal{G}_3 + \frac{1}{2}\left(\mathcal{G}_{1,t} - \mathcal{G}_{2,t} - 2\mathcal{T}_2\right)\cos^2\mathcal{G}_3\,,\nonumber\\
\mathcal{C}_8 = \frac{1}{2}e^{-\mathcal{G}_1}\left[(\mathcal{Q}_2 - 2\mathcal{Q}_7)\sin\mathcal{G}_3 - \mathcal{Q}_6e^{\mathcal{G}_2}\cos\mathcal{G}_3\right] + \frac{1}{4}\left[\left(2\mathcal{T}_2 - \mathcal{G}_{1,t} + \mathcal{G}_{2,t}\right)e^{-\mathcal{G}_2} + \mathcal{G}_{3,r}\right]\sin(2\mathcal{G}_3)\nonumber\\
+ \frac{1}{4}\mathcal{G}_{1,r}(3 - \cos(2\mathcal{G}_3)) - \frac{1}{2}\mathcal{G}_{2,r}\cos^2\mathcal{G}_3 + \frac{1}{2}\left(2\mathcal{T}_1 + \mathcal{G}_{3,t}e^{-\mathcal{G}_2}\right)\sin^2\mathcal{G}_3\,,\nonumber\\
\mathcal{C}_9 = \frac{1}{2}e^{-\mathcal{G}_1}\left\{\left[\mathcal{Q}_8 - 2\mathcal{Q}_{10} + \left(2\mathcal{T}_6 - \mathcal{G}_{4,r}\right)e^{\mathcal{G}_4}\right]\sin\mathcal{G}_3 - \left[\mathcal{Q}_4 - 2\mathcal{Q}_9 + \left(2\mathcal{T}_5 - \mathcal{G}_{4,t}\right)e^{\mathcal{G}_4}\right]e^{\mathcal{G}_2}\cos\mathcal{G}_3\right\}\,,\nonumber\\
\mathcal{C}_{10} = \frac{1}{2}e^{-\mathcal{G}_1}\left\{\left[\mathcal{Q}_4 - 2\mathcal{Q}_9 + \left(2\mathcal{T}_5 - \mathcal{G}_{4,t}\right)e^{\mathcal{G}_4}\right]\sin\mathcal{G}_3 + \left[\mathcal{Q}_8 - 2\mathcal{Q}_{10} + \left(2\mathcal{T}_6 - \mathcal{G}_{4,r}\right)e^{\mathcal{G}_4}\right]e^{-\mathcal{G}_2}\cos\mathcal{G}_3\right\}\,,\nonumber\\
\mathcal{C}_{15} = -\frac{1}{2}e^{-\mathcal{G}_4}\left[2\mathcal{Q}_{11} + e^{\mathcal{G}_1}\left(\mathcal{T}_4\sin\mathcal{G}_3 - \mathcal{T}_3e^{\mathcal{G}_2}\cos\mathcal{G}_3\right)\right]\,, \quad
\mathcal{C}_{13} = \frac{1}{2}\left(\mathcal{G}_{4,t} - e^{-\mathcal{G}_4}\mathcal{Q}_4\right)\,,\nonumber\\
\mathcal{C}_{16} = -\frac{1}{2}e^{-\mathcal{G}_4}\left[2\mathcal{Q}_{12} + e^{\mathcal{G}_1}\left(\mathcal{T}_3\sin\mathcal{G}_3 + \mathcal{T}_4e^{-\mathcal{G}_2}\cos\mathcal{G}_3\right)\right]\,, \quad
\mathcal{C}_{14} = \frac{1}{2}\left(\mathcal{G}_{4,r} - e^{-\mathcal{G}_4}\mathcal{Q}_8\right)\,,\nonumber\\
\mathcal{C}_{11} = \mathcal{C}_{13} - \mathcal{T}_5\,, \quad
\mathcal{C}_{12} = \mathcal{C}_{14} - \mathcal{T}_6\,, \quad
\mathcal{C}_2 = \mathcal{C}_3 - \mathcal{T}_1\,, \quad
\mathcal{C}_{19} = \frac{1}{2}\mathcal{T}_3\,,\nonumber\\
\mathcal{C}_{17} = \mathcal{C}_{15} - \mathcal{T}_7\,, \quad
\mathcal{C}_{18} = \mathcal{C}_{16} - \mathcal{T}_8\,, \quad
\mathcal{C}_7 = \mathcal{C}_6 + \mathcal{T}_2\,, \quad
\mathcal{C}_{20} = \frac{1}{2}\mathcal{T}_4\,.
\end{gather}
The advantage of the new parametrization becomes apparent when one derives the torsion and nonmetricity of the connection. In terms of the new variables, the non-vanishing and independent components of the torsion are given by
\begin{align}
T^t{}_{tr} &= \mathcal{T}_1\,, &
T^t{}_{\vartheta\varphi} &= \mathcal{T}_3\sin\vartheta\,, &
T^{\vartheta}{}_{t\vartheta} &= T^{\varphi}{}_{t\varphi} = \mathcal{T}_5\,, &
T^{\vartheta}{}_{t\varphi} &= \mathcal{T}_7\sin\vartheta\,, &
T^{\varphi}{}_{t\vartheta} &= -\frac{\mathcal{T}_7}{\sin\vartheta}\,,\nonumber\\
T^r{}_{tr} &= \mathcal{T}_2\,, &
T^r{}_{\vartheta\varphi} &= \mathcal{T}_4\sin\vartheta\,, &
T^{\vartheta}{}_{r\vartheta} &= T^{\varphi}{}_{r\varphi} = \mathcal{T}_6\,, &
T^{\vartheta}{}_{r\varphi} &= \mathcal{T}_8\sin\vartheta\,, &
T^{\varphi}{}_{r\vartheta} &= -\frac{\mathcal{T}_8}{\sin\vartheta}\,,\label{eq:torsiont}
\end{align}
while the non-vanishing, independent components of the nonmetricity are given by
\begin{gather}
Q_{ttt} = \mathcal{Q}_1\,, \quad
Q_{trr} = \mathcal{Q}_2\,, \quad
Q_{ttr} = \mathcal{Q}_3\,, \quad
Q_{t\vartheta\vartheta} = \mathcal{Q}_4\,, \quad
Q_{t\varphi\varphi} = \mathcal{Q}_4\sin^2\vartheta\,, \quad
Q_{\vartheta t\vartheta} = \mathcal{Q}_9\,,\nonumber\\
Q_{rtt} = \mathcal{Q}_5\,, \quad
Q_{rrr} = \mathcal{Q}_6\,, \quad
Q_{rtr} = \mathcal{Q}_7\,, \quad
Q_{r\vartheta\vartheta} = \mathcal{Q}_8\,, \quad
Q_{r\varphi\varphi} = \mathcal{Q}_8\sin^2\vartheta\,, \quad
Q_{\vartheta r\vartheta} = \mathcal{Q}_{10}\,,\label{eq:nonmetq}\\
Q_{\varphi t\varphi} = \mathcal{Q}_9\sin^2\vartheta\,, \quad
Q_{\varphi r\varphi} = \mathcal{Q}_{10}\sin^2\vartheta\,, \quad
Q_{\vartheta t\varphi} = -Q_{\varphi t\vartheta} = \mathcal{Q}_{11}\sin\vartheta\,, \quad
Q_{\vartheta r\varphi} = -Q_{\varphi r\vartheta} = \mathcal{Q}_{12}\sin\vartheta\,.\nonumber
\end{gather}
We see that the newly introduced parameters simply parametrize the components of the torsion and nonmetricity. This new parametrization turns out to be more suitable in order to express the components of the contortion and the disformation than the parametrizations we introduced in the previous section. For the contortion it is most useful to display the components with lower indices, in order to exploit the antisymmetry \(K_{\mu\nu\rho} = -K_{\nu\mu\rho}\) in the first two indices, and thus to reduce the number of independent components. We find that the contortion is given by
\begin{align}
K_{t\varphi\vartheta} &= -K_{t\vartheta\varphi} = \frac{1}{2}e^{\mathcal{G}_1}\left(\mathcal{T}_4\sin\mathcal{G}_3 - \mathcal{T}_3e^{\mathcal{G}_2}\cos\mathcal{G}_3\right)\sin\vartheta\,, &
K_{trt} &= e^{\mathcal{G}_1}\left(\mathcal{T}_2\sin\mathcal{G}_3 - \mathcal{T}_1e^{\mathcal{G}_2}\cos\mathcal{G}_3\right)\,,\nonumber\\
K_{r\varphi\vartheta} &= -K_{r\vartheta\varphi} = \frac{1}{2}e^{\mathcal{G}_1}\left(\mathcal{T}_3\sin\mathcal{G}_3 + \mathcal{T}_4e^{-\mathcal{G}_2}\cos\mathcal{G}_3\right)\sin\vartheta\,, &
K_{trr} &= e^{\mathcal{G}_1}\left(\mathcal{T}_1\sin\mathcal{G}_3 + \mathcal{T}_2e^{-\mathcal{G}_2}\cos\mathcal{G}_3\right)\,,\nonumber\\
K_{\vartheta\varphi t} &= \frac{1}{2}\left[2\mathcal{T}_7e^{\mathcal{G}_4} + \left(\mathcal{T}_4\sin\mathcal{G}_3 - \mathcal{T}_3e^{\mathcal{G}_2}\cos\mathcal{G}_3\right)e^{\mathcal{G}_1}\right]\sin\vartheta\,, &
K_{t\vartheta\vartheta} &= \frac{K_{t\varphi\varphi}}{\sin^2\vartheta} = e^{\mathcal{G}_4}\mathcal{T}_5\,,\nonumber\\
K_{\vartheta\varphi r} &= \frac{1}{2}\left[2\mathcal{T}_8e^{\mathcal{G}_4} + \left(\mathcal{T}_3\sin\mathcal{G}_3 + \mathcal{T}_4e^{-\mathcal{G}_2}\cos\mathcal{G}_3\right)e^{\mathcal{G}_1}\right]\sin\vartheta\,, &
K_{r\vartheta\vartheta} &= \frac{K_{r\varphi\varphi}}{\sin^2\vartheta} = e^{\mathcal{G}_4}\mathcal{T}_6\,.\label{eq:contortiont}
\end{align}
For easier comparison, we display the disformation using the same index positions. Here the non-vanishing and independent components are given by
\begin{align}
L_{ttt} &= -\frac{1}{2}\mathcal{Q}_1\,, &
L_{trr} &= \frac{1}{2}\mathcal{Q}_2 - \mathcal{Q}_7\,, &
L_{t\vartheta\vartheta} &= \frac{L_{t\varphi\varphi}}{\sin^2\vartheta} = \frac{1}{2}\mathcal{Q}_4 - \mathcal{Q}_9\,,\nonumber\\
L_{rrr} &= -\frac{1}{2}\mathcal{Q}_6\,, &
L_{rtt} &= \frac{1}{2}\mathcal{Q}_5 - \mathcal{Q}_3\,, &
L_{r\vartheta\vartheta} &= \frac{L_{r\varphi\varphi}}{\sin^2\vartheta} = \frac{1}{2}\mathcal{Q}_8 - \mathcal{Q}_{10}\,,\nonumber\\
L_{ttr} &= -\frac{1}{2}\mathcal{Q}_5\,, &
L_{\vartheta t\varphi} &= -L_{\varphi t\vartheta} = \mathcal{Q}_{11}\sin\vartheta\,, &
L_{\vartheta t\vartheta} &= \frac{L_{\varphi t\varphi}}{\sin^2\theta} = -\frac{1}{2}\mathcal{Q}_4\,,\nonumber\\
L_{rtr} &= -\frac{1}{2}\mathcal{Q}_2\,, &
L_{\vartheta r\varphi} &= -L_{\varphi r\vartheta} = \mathcal{Q}_{12}\sin\vartheta\,, &
L_{\vartheta r\vartheta} &= \frac{L_{\varphi r\varphi}}{\sin^2\theta} = -\frac{1}{2}\mathcal{Q}_8\,.\label{eq:disformationq}
\end{align}
One now easily checks that, together with the components
\begin{align}
\lc{\Gamma}_{ttt} &= \frac{1}{2}e^{\mathcal{G}_1 + \mathcal{G}_2}\left[\mathcal{G}_{3,t}\sin\mathcal{G}_3 - (\mathcal{G}_{1,t} + \mathcal{G}_{2,t})\cos\mathcal{G}_3\right]\,,\nonumber\\
\lc{\Gamma}_{ttr} &= \frac{1}{2}e^{\mathcal{G}_1 + \mathcal{G}_2}\left[\mathcal{G}_{3,r}\sin\mathcal{G}_3 - (\mathcal{G}_{1,r} + \mathcal{G}_{2,r})\cos\mathcal{G}_3\right]\,,\nonumber\\
\lc{\Gamma}_{trr} &= \frac{1}{2}e^{\mathcal{G}_1}\left\{\left[2\mathcal{G}_{3,r} - \left(\mathcal{G}_{1,t} - \mathcal{G}_{2,t}\right)e^{-\mathcal{G}_2}\right]\cos\mathcal{G}_3 + \left(2\mathcal{G}_{1,r} + \mathcal{G}_{3,t}e^{-\mathcal{G}_2}\right)\sin\mathcal{G}_3\right\}\,,\nonumber\\
\lc{\Gamma}_{rtt} &= \frac{1}{2}e^{\mathcal{G}_1}\left\{\left[2\mathcal{G}_{3,t} + \left(\mathcal{G}_{1,r} + \mathcal{G}_{2,r}\right)e^{\mathcal{G}_2}\right]\cos\mathcal{G}_3 + \left(2\mathcal{G}_{1,t} - \mathcal{G}_{3,r}e^{\mathcal{G}_2}\right)\sin\mathcal{G}_3\right\}\,,\nonumber\\
\lc{\Gamma}_{rtr} &= -\frac{1}{2}e^{\mathcal{G}_1 - \mathcal{G}_2}\left[\mathcal{G}_{3,t}\sin\mathcal{G}_3 - (\mathcal{G}_{1,t} - \mathcal{G}_{2,t})\cos\mathcal{G}_3\right]\,,\nonumber\\
\lc{\Gamma}_{rrr} &= -\frac{1}{2}e^{\mathcal{G}_1 - \mathcal{G}_2}\left[\mathcal{G}_{3,r}\sin\mathcal{G}_3 - (\mathcal{G}_{1,r} - \mathcal{G}_{2,r})\cos\mathcal{G}_3\right]\,,\nonumber\\
\lc{\Gamma}_{t\vartheta\vartheta} &= -\lc{\Gamma}_{\vartheta t\vartheta} = \frac{\lc{\Gamma}_{t\varphi\varphi}}{\sin^2\vartheta} = -\frac{\lc{\Gamma}_{\varphi t\varphi}}{\sin^2\vartheta} = -\frac{1}{2}e^{\mathcal{G}_4}\mathcal{G}_{4,t}\,,\nonumber\\
\lc{\Gamma}_{r\vartheta\vartheta} &= -\lc{\Gamma}_{\vartheta r\vartheta} = \frac{\lc{\Gamma}_{r\varphi\varphi}}{\sin^2\vartheta} = -\frac{\lc{\Gamma}_{\varphi r\varphi}}{\sin^2\vartheta} = -\frac{1}{2}e^{\mathcal{G}_4}\mathcal{G}_{4,r}\,,\nonumber\\
\lc{\Gamma}_{\varphi\vartheta\varphi} &= -\lc{\Gamma}_{\vartheta\varphi\varphi} = e^{\mathcal{G}_4}\cos\vartheta\sin\vartheta\label{eq:levicivita}
\end{align}
of the Levi-Civita connection, the relation~\eqref{eq:conndecomp} is indeed satisfied.

\subsection{Curvature}\label{ssec:curv}
Finally, we calculate the curvature of the general spherically symmetric connection. It is instructive to divide its components into two classes. First, note that the 6 components
\begin{align}
R^t{}_{t\vartheta\varphi} &= 2(\mathcal{C}_{11}\mathcal{C}_{19} - \mathcal{C}_9\mathcal{C}_{15})\sin\vartheta\,,\nonumber\\
R^t{}_{r\vartheta\varphi} &= 2(\mathcal{C}_{12}\mathcal{C}_{19} - \mathcal{C}_9\mathcal{C}_{16})\sin\vartheta\,,\nonumber\\
R^r{}_{t\vartheta\varphi} &= 2(\mathcal{C}_{11}\mathcal{C}_{20} - \mathcal{C}_{10}\mathcal{C}_{15})\sin\vartheta\,,\nonumber\\
R^r{}_{r\vartheta\varphi} &= 2(\mathcal{C}_{12}\mathcal{C}_{20} - \mathcal{C}_{10}\mathcal{C}_{16})\sin\vartheta\,,\nonumber\\
R^{\vartheta}{}_{\vartheta\vartheta\varphi} &= R^{\varphi}{}_{\varphi\vartheta\varphi} = \left(\mathcal{C}_9\mathcal{C}_{15} + \mathcal{C}_{10}\mathcal{C}_{16} - \mathcal{C}_{11}\mathcal{C}_{19} - \mathcal{C}_{12}\mathcal{C}_{20}\right)\sin\vartheta\,,\nonumber\\
R^{\vartheta}{}_{\varphi\vartheta\varphi} &= -R^{\varphi}{}_{\vartheta\vartheta\varphi}\sin^2\vartheta = \left(1 + \mathcal{C}_9\mathcal{C}_{11} + \mathcal{C}_{10}\mathcal{C}_{12} + \mathcal{C}_{15}\mathcal{C}_{19} + \mathcal{C}_{16}\mathcal{C}_{20}\right)\sin^2\vartheta\label{eq:curv1}
\end{align}
depend only algebraically on 8 of the parameter functions \(\mathcal{C}_1, \ldots, \mathcal{C}_{20}\). The vanishing of these components yields 5 independent equations, which determine a hyperbolic submanifold of the total parameter space; any flat, i.e., curvature-free connection, as we encounter in section~\ref{ssec:nor}, lies within this submanifold. The remaining components depend also on the derivatives of the parameter functions. In particular, we find the equations
\begin{align}
R^{\vartheta}{}_{\vartheta tr} &= R^{\varphi}{}_{\varphi tr} = \mathcal{C}_{14,t} - \mathcal{C}_{13,r}\,,\nonumber\\
R^{\vartheta}{}_{\varphi tr} &= -R^{\varphi}{}_{\vartheta tr}\sin^2\vartheta = -\left(\mathcal{C}_{18,t} - \mathcal{C}_{17,r}\right)\sin\vartheta\,,\label{eq:curv2}
\end{align}
as well as
\begin{align}
R^t{}_{ttr} &= \mathcal{C}_{2,t} - \mathcal{C}_{1,r} + \mathcal{C}_3\mathcal{C}_6 - \mathcal{C}_4\mathcal{C}_5\,,\nonumber\\
R^t{}_{rtr} &= \mathcal{C}_{4,t} - \mathcal{C}_{3,r} + \mathcal{C}_4(\mathcal{C}_1 - \mathcal{C}_7) - \mathcal{C}_3(\mathcal{C}_2 - \mathcal{C}_8)\,,\nonumber\\
R^r{}_{ttr} &= \mathcal{C}_{6,t} - \mathcal{C}_{5,r} + \mathcal{C}_6(\mathcal{C}_7 - \mathcal{C}_1) - \mathcal{C}_5(\mathcal{C}_8 - \mathcal{C}_2)\,,\nonumber\\
R^r{}_{rtr} &= \mathcal{C}_{8,t} - \mathcal{C}_{7,r} + \mathcal{C}_4\mathcal{C}_5 - \mathcal{C}_3\mathcal{C}_6\,,\label{eq:curv3}
\end{align}
which can be understood as defining integrability conditions on the remaining 12 parameter functions. Finally, the components
\begin{align}
R^t{}_{\vartheta t\vartheta} &= \frac{R^t{}_{\varphi t\varphi}}{\sin^2\vartheta} = \mathcal{C}_{9,t} + \mathcal{C}_3\mathcal{C}_{10} - \mathcal{C}_{17}\mathcal{C}_{19} + \mathcal{C}_9(\mathcal{C}_1 - \mathcal{C}_{13})\,,\nonumber\\
R^t{}_{\vartheta r\vartheta} &= \frac{R^t{}_{\varphi r\varphi}}{\sin^2\vartheta} = \mathcal{C}_{9,r} + \mathcal{C}_4\mathcal{C}_{10} - \mathcal{C}_{18}\mathcal{C}_{19} + \mathcal{C}_9(\mathcal{C}_2 - \mathcal{C}_{14})\,,\nonumber\\
R^r{}_{\vartheta t\vartheta} &= \frac{R^r{}_{\varphi t\varphi}}{\sin^2\vartheta} = \mathcal{C}_{10,t} + \mathcal{C}_5\mathcal{C}_9 - \mathcal{C}_{17}\mathcal{C}_{20} + \mathcal{C}_{10}(\mathcal{C}_7 - \mathcal{C}_{13})\,,\nonumber\\
R^r{}_{\vartheta r\vartheta} &= \frac{R^r{}_{\varphi r\varphi}}{\sin^2\vartheta} = \mathcal{C}_{10,r} + \mathcal{C}_6\mathcal{C}_9 - \mathcal{C}_{18}\mathcal{C}_{20} + \mathcal{C}_{10}(\mathcal{C}_8 - \mathcal{C}_{14})\,,\nonumber\\
R^{\vartheta}{}_{tt\vartheta} &= R^{\varphi}{}_{tt\varphi} = \mathcal{C}_{11,t} - \mathcal{C}_5\mathcal{C}_{12} - \mathcal{C}_{15}\mathcal{C}_{17} + \mathcal{C}_{11}(\mathcal{C}_{13} - \mathcal{C}_1)\,,\nonumber\\
R^{\vartheta}{}_{tr\vartheta} &= R^{\varphi}{}_{tr\varphi} = \mathcal{C}_{11,r} - \mathcal{C}_6\mathcal{C}_{12} - \mathcal{C}_{15}\mathcal{C}_{18} + \mathcal{C}_{11}(\mathcal{C}_{14} - \mathcal{C}_2)\,,\nonumber\\
R^{\vartheta}{}_{rt\vartheta} &= R^{\varphi}{}_{rt\varphi} = \mathcal{C}_{12,t} - \mathcal{C}_3\mathcal{C}_{11} - \mathcal{C}_{16}\mathcal{C}_{17} + \mathcal{C}_{12}(\mathcal{C}_{13} - \mathcal{C}_7)\,,\nonumber\\
R^{\vartheta}{}_{rr\vartheta} &= R^{\varphi}{}_{rt\varphi} = \mathcal{C}_{12,r} - \mathcal{C}_4\mathcal{C}_{11} - \mathcal{C}_{16}\mathcal{C}_{18} + \mathcal{C}_{12}(\mathcal{C}_{14} - \mathcal{C}_8)\,,\nonumber\\
R^t{}_{\vartheta t\varphi} &= -R^t{}_{\varphi t\vartheta} = -\left[\mathcal{C}_{19,t} + \mathcal{C}_9\mathcal{C}_{17} + \mathcal{C}_3\mathcal{C}_{20} + \mathcal{C}_{19}(\mathcal{C}_1 - \mathcal{C}_{13})\right]\sin\vartheta\,,\nonumber\\
R^t{}_{\vartheta r\varphi} &= -R^t{}_{\varphi r\vartheta} = -\left[\mathcal{C}_{19,r} + \mathcal{C}_9\mathcal{C}_{18} + \mathcal{C}_4\mathcal{C}_{20} + \mathcal{C}_{19}(\mathcal{C}_2 - \mathcal{C}_{14})\right]\sin\vartheta\,,\nonumber\\
R^r{}_{\vartheta t\varphi} &= -R^r{}_{\varphi t\vartheta} = -\left[\mathcal{C}_{20,t} + \mathcal{C}_{10}\mathcal{C}_{17} + \mathcal{C}_5\mathcal{C}_{19} + \mathcal{C}_{20}(\mathcal{C}_7 - \mathcal{C}_{13})\right]\sin\vartheta\,,\nonumber\\
R^r{}_{\vartheta r\varphi} &= -R^r{}_{\varphi r\vartheta} = -\left[\mathcal{C}_{20,r} + \mathcal{C}_{10}\mathcal{C}_{18} + \mathcal{C}_6\mathcal{C}_{19} + \mathcal{C}_{20}(\mathcal{C}_8 - \mathcal{C}_{14})\right]\sin\vartheta\,,\nonumber\\
R^{\vartheta}{}_{tt\varphi} &= -R^{\varphi}{}_{tt\vartheta}\sin^2\vartheta = -\left[\mathcal{C}_{15,t} + \mathcal{C}_{11}\mathcal{C}_{17} - \mathcal{C}_5\mathcal{C}_{16} + \mathcal{C}_{15}(\mathcal{C}_{13} - \mathcal{C}_1)\right]\sin\vartheta\,,\nonumber\\
R^{\vartheta}{}_{tr\varphi} &= -R^{\varphi}{}_{tr\vartheta}\sin^2\vartheta = -\left[\mathcal{C}_{15,r} + \mathcal{C}_{11}\mathcal{C}_{18} - \mathcal{C}_6\mathcal{C}_{16} + \mathcal{C}_{15}(\mathcal{C}_{14} - \mathcal{C}_2)\right]\sin\vartheta\,,\nonumber\\
R^{\vartheta}{}_{rt\varphi} &= -R^{\varphi}{}_{rt\vartheta}\sin^2\vartheta = -\left[\mathcal{C}_{16,t} + \mathcal{C}_{12}\mathcal{C}_{17} - \mathcal{C}_3\mathcal{C}_{15} + \mathcal{C}_{16}(\mathcal{C}_{13} - \mathcal{C}_7)\right]\sin\vartheta\,,\nonumber\\
R^{\vartheta}{}_{rr\varphi} &= -R^{\varphi}{}_{rr\vartheta}\sin^2\vartheta = -\left[\mathcal{C}_{16,r} + \mathcal{C}_{12}\mathcal{C}_{18} - \mathcal{C}_4\mathcal{C}_{15} + \mathcal{C}_{16}(\mathcal{C}_{14} - \mathcal{C}_8)\right]\sin\vartheta\,,\label{eq:curv4}
\end{align}
which depend on derivatives of those 8 parameter functions which are restricted by the algebraic equations~\eqref{eq:curv1}, intertwine these two sets of parameter functions. This separation of equations may be exploited to determine the most general flat connection; however, we will use a different approach, which we display in section~\ref{ssec:nor}.

\section{Special cases}\label{sec:special}
In the previous section we have considered a fully general connection, which may have torsion, nonmetricity and curvature. We now turn our focus to more restricted connections, by imposing that one or more of these properties vanish. This will lead us to the cases \(T = 0\) in section~\ref{ssec:not}, \(Q = 0\) in section~\ref{ssec:noq}, \(T = Q = 0\) in section~\ref{ssec:notq}, \(R = 0\) in section~\ref{ssec:nor}, \(R = T = 0\) in section~\ref{ssec:nort}, \(R = Q = 0\) in section~\ref{ssec:norq} and finally \(R = T = Q = 0\) in section~\ref{ssec:nortq}.

\subsection{Torsion-free: $T = 0$}\label{ssec:not}
In order to determine the most general torsion-free connection with spherical symmetry, it is most practical to use the parametrization we introduced in section~\ref{ssec:conndecomp}. In this parametrization one finds immediately that the torsion~\eqref{eq:torsionc}, which reduces to the form~\eqref{eq:torsiont}, vanishes if and only if \(\mathcal{T}_1 = \ldots = \mathcal{T}_8 = 0\). The most general torsion-free metric-affine geometry we are looking for is thus determined by the parameter functions \(\mathcal{G}_1, \ldots, \mathcal{G}_4\) determining the metric, as well as the parameter functions \(\mathcal{Q}_1, \ldots, \mathcal{Q}_{12}\) determining the nonmetricity. Note that the connection can most conveniently be expressed through the relation~\eqref{eq:conndecomp}, with the disformation~\eqref{eq:disformationq} and the Levi-Civita connection~\eqref{eq:levicivita}.

\subsection{Metric-compatible: $Q = 0$}\label{ssec:noq}
The converse case, compared to the previous one, is a general metric-compatible connection, while allowing for non-vanishing curvature and torsion, i.e., a Riemann-Cartan geometry. As in the torsion-free case, the parametrization introduced in section~\ref{ssec:conndecomp} immediately yields the desired result, in this case by choosing the parameters \(\mathcal{Q}_1 = \ldots = \mathcal{Q}_{12} = 0\). The resulting metric-affine geometry is thus parametrized by the parameter functions \(\mathcal{G}_1, \ldots, \mathcal{G}_4\) determining the metric, as well as the parameter functions \(\mathcal{T}_1, \ldots, \mathcal{T}_8\) determining the torsion. Also in this case the connection is expressed through the relation~\eqref{eq:conndecomp}, now with the contortion~\eqref{eq:contortiont} and the Levi-Civita connection~\eqref{eq:levicivita}.

\subsection{Torsion-free metric-compatible: $T = Q = 0$}\label{ssec:notq}
For the sake of completeness we mention that by choosing the parameter functions \(\mathcal{T}_1 = \ldots = \mathcal{T}_8 = \mathcal{Q}_1 = \ldots = \mathcal{Q}_{20} = 0\) one obtains the unique metric-compatible and torsion-free connection, which is, of course, the Levi-Civita connection~\eqref{eq:levicivita}.

\subsection{Flat: $R = 0$}\label{ssec:nor}
There are different possibilities to derive flat, symmetric metric-affine geometries, i.e., symmetric metric-affine geometries with vanishing curvature. The most straightforward approach is to consider the general spherically symmetric connection derived in section~\ref{ssec:conn}, and to impose that its curvature vanishes. This results in a number of differential equations which are quadratic in the unknowns to be solved for, which may be involved, depending on the degree of symmetry imposed. Another strategy is to realize that the existence of a flat connection implies (on a simply-connected manifold) the existence of a global coframe \(\Theta^a{}_{\mu}\), which is in general different from the metric coframe \(\theta^a{}_{\mu}\), and which may be constructed by choosing the coframe in a single spacetime point \(x\), and then using the path-independent parallel transport defined by the flat connection to obtain the coframe in any other spacetime point. It follows from this construction that the tetrad components are covariantly constant,
\begin{equation}
0 = \nabla_{\mu}\Theta^a{}_{\nu} = \partial_{\mu}\Theta^a{}_{\nu} - \Gamma^{\rho}{}_{\nu\mu}\Theta^a{}_{\rho}\,,
\end{equation}
so that the connection coefficients are given by the Weitzenböck connection
\begin{equation}\label{eq:wbconn}
\Gamma^{\mu}{}_{\nu\rho} = E_a{}^{\mu}\partial_{\rho}\Theta^a{}_{\nu}\,,
\end{equation}
where \(E_a{}^{\mu}\) is the inverse tetrad satisfying \(\Theta^a{}_{\mu}E_b{}^{\mu} = \delta^a_b\) and \(\Theta^a{}_{\mu}E_a{}^{\nu} = \delta_{\mu}^{\nu}\). One may then obtain the symmetric coframe, and hence the flat symmetric connection, by inserting the Weitzenböck connection in the Lie derivative~\eqref{eq:affsymcondi} and solving the resulting equations for the tetrad. Taking into account that the tetrad transforms as a one-form, so that its Lie derivative with respect to a symmetry generator \(X_{\xi}\) is given by
\begin{equation}
(\mathcal{L}_{X_{\xi}}\Theta)^a{}_{\mu} = X_{\xi}^{\nu}\partial_{\nu}\Theta^a{}_{\mu} + \partial_{\mu}X_{\xi}^{\nu}\Theta^a{}_{\nu}\,,
\end{equation}
one finds that \((\mathcal{L}_{X_{\xi}}\Gamma)^{\mu}{}_{\nu\rho} = 0\), i.e., the Weitzenböck connection obeys the symmetry, if and only if
\begin{equation}
0 = \nabla_{\mu}(\mathcal{L}_{X_{\xi}}\Theta)^a{}_{\nu} = \partial_{\mu}(\mathcal{L}_{X_{\xi}}\Theta)^a{}_{\nu} - \Gamma^{\rho}{}_{\nu\mu}(\mathcal{L}_{X_{\xi}}\Theta)^a{}_{\rho}\,,
\end{equation}
where the connection coefficients are given by the definition~\eqref{eq:wbconn}. Note that by construction, the Weitzenböck connection is necessarily a metric connection, though not with respect to the independent metric \(g_{\mu\nu}\), but with respect to the metric \(\tilde{g}_{\mu\nu} = \eta_{ab}\Theta^a{}_{\mu}\Theta^b{}_{\nu}\) defined by the parallely transported tetrad.

Observe that the metric \(\tilde{g}_{\mu\nu}\) is not fully defined by the connection alone, but also depends on the choice on the tetrad \(\Theta^a{}_{\mu}(x)\) at the initial point \(x\) of the construction above. The Weitzenböck connection does not depend on the initial tetrad, since any constant linear transformation cancels in its definition~\eqref{eq:wbconn}. Hence, we are free to make a convenient choice. Using the fact that the symmetry group generating the spherical symmetry is \(\mathrm{SO}(3)\), we may distinguish two cases:
\begin{enumerate}
\item
The action \(\Theta^a{}_{\mu}(x) \mapsto (\boldsymbol{\Lambda}_u^{-1})^a{}_b(x)\Theta^b{}_{\mu}(x)\) is trivial. This case is topologically excluded, since in this case the parallel transport of the tetrad along the orbit of the symmetry group, which is topologically a sphere \(S^2\), would yield a global frame on the sphere. However, this is impossible, since the sphere is not parallelizable. See~\cite{Hohmann:2019nat} for a detailed derivation of this contradiction.

\item
If the action of the symmetry group on the tetrad \(\Theta^a{}_{\mu}(x)\) is non-trivial, one can use the properties of \(G = \mathrm{SO}(3)\) to realize that the image of \(G\) under the map \(u \mapsto \boldsymbol{\Lambda}_u(x) \in \mathrm{GL}(4)\) is again isomorphic to \(\mathrm{SO}(3)\) itself. Hence, one may always find a tetrad such that its temporal component \(\Theta^0{}_{\mu}(x)\) is invariant under the group action, while its spatial components span an orthonormal basis for the rotation group. It then follows that \(\boldsymbol{\Lambda}_u(x) \in \mathrm{SO}(3) \subset SO(1,3)\) for all \(u \in G\).
\end{enumerate}
Finally, note that the symmetry of the Weitzenböck connection, whose spin connection vanishes by definition, implies that \(\boldsymbol{\Lambda}_u\) as defined above does not depend on the spacetime point, \(\partial_{\mu}\boldsymbol{\Lambda}_u = 0\), and so globally defines an element of the Lorentz group. This means that the metric \(\tilde{g}_{\mu\nu}\) defined by this tetrad is symmetric under the action of the symmetry group, following the derivation in section~\ref{ssec:tetsym}.

In summary, we thus find that the tetrad \(\Theta^a{}_{\mu}\) defines a metric \(\tilde{g}_{\mu\nu}\) and a metric-compatible, flat Weitzenböck connection, both of which adhere to the spherical symmetry. The most general tetrad which satisfies these conditions depends on 6 free functions \(\mathcal{F}_1, \ldots, \mathcal{F}_6\) and takes the form~\cite{Hohmann:2019nat}
\begin{subequations}\label{eq:sphertetradwb}
\begin{align}
\Theta^0 &= \mathcal{F}_1\cosh\mathcal{F}_3\dd t + \mathcal{F}_2\sinh\mathcal{F}_4\dd r\,,\\
\Theta^1 &= \sin\vartheta\cos\varphi(\mathcal{F}_1\sinh\mathcal{F}_3\dd t + \mathcal{F}_2\cosh\mathcal{F}_4\dd r)\nonumber\\
&\phantom{=}+ \mathcal{F}_5\left[(\cos\mathcal{F}_6\cos\vartheta\cos\varphi - \sin\mathcal{F}_6\sin\varphi)\dd\vartheta - \sin\vartheta(\cos\mathcal{F}_6\sin\varphi + \sin\mathcal{F}_6\cos\vartheta\cos\varphi)\dd\varphi\right]\,,\\
\Theta^2 &= \sin\vartheta\sin\varphi(\mathcal{F}_1\sinh\mathcal{F}_3\dd t + \mathcal{F}_2\cosh\mathcal{F}_4\dd r)\nonumber\\
&\phantom{=}+ \mathcal{F}_5\left[(\cos\mathcal{F}_6\cos\vartheta\sin\varphi + \sin\mathcal{F}_6\cos\varphi)\dd\vartheta + \sin\vartheta(\cos\mathcal{F}_6\cos\varphi - \sin\mathcal{F}_6\cos\vartheta\sin\varphi)\dd\varphi\right]\,,\\
\Theta^3 &= \cos\vartheta(\mathcal{F}_1\sinh\mathcal{F}_3\dd t + \mathcal{F}_2\cosh\mathcal{F}_4\dd r) + \mathcal{F}_5\left[-\cos\mathcal{F}_6\sin\vartheta\dd\vartheta + \sin\mathcal{F}_6\sin^2\vartheta\dd\varphi\right]\,.
\end{align}
\end{subequations}
The corresponding Weitzenböck connection then takes the general form~\eqref{eq:sphergamma}, where the parameter functions \(\mathcal{C}_1, \ldots, \mathcal{C}_{20}\) are given by
\begin{align}
\mathcal{C}_1 &= \frac{\mathcal{F}_{1,t}}{\mathcal{F}_1} + \mathcal{F}_{3,t}\tanh(\mathcal{F}_3 - \mathcal{F}_4)\,, &
\mathcal{C}_7 &= \frac{\mathcal{F}_{2,t}}{\mathcal{F}_2} - \mathcal{F}_{4,t}\tanh(\mathcal{F}_3 - \mathcal{F}_4)\,,\nonumber\\
\mathcal{C}_2 &= \frac{\mathcal{F}_{1,r}}{\mathcal{F}_1} + \mathcal{F}_{3,r}\tanh(\mathcal{F}_3 - \mathcal{F}_4)\,, &
\mathcal{C}_8 &= \frac{\mathcal{F}_{2,r}}{\mathcal{F}_2} - \mathcal{F}_{4,r}\tanh(\mathcal{F}_3 - \mathcal{F}_4)\,,\nonumber\\
\mathcal{C}_3 &= \frac{\mathcal{F}_2\mathcal{F}_{4,t}}{\mathcal{F}_1\cosh(\mathcal{F}_3 - \mathcal{F}_4)}\,, &
\mathcal{C}_5 &= \frac{\mathcal{F}_1\mathcal{F}_{3,t}}{\mathcal{F}_2\cosh(\mathcal{F}_3 - \mathcal{F}_4)}\,,\nonumber\\
\mathcal{C}_4 &= \frac{\mathcal{F}_2\mathcal{F}_{4,r}}{\mathcal{F}_1\cosh(\mathcal{F}_3 - \mathcal{F}_4)}\,, &
\mathcal{C}_6 &= \frac{\mathcal{F}_1\mathcal{F}_{3,r}}{\mathcal{F}_2\cosh(\mathcal{F}_3 - \mathcal{F}_4)}\,,\nonumber\\
\mathcal{C}_9 &= \frac{\mathcal{F}_5\sinh\mathcal{F}_4\cos\mathcal{F}_6}{\mathcal{F}_1\cosh(\mathcal{F}_3 - \mathcal{F}_4)}\,, &
\mathcal{C}_{10} &= -\frac{\mathcal{F}_5\cosh\mathcal{F}_3\cos\mathcal{F}_6}{\mathcal{F}_2\cosh(\mathcal{F}_3 - \mathcal{F}_4)}\,,\nonumber\\
\mathcal{C}_{20} &= \frac{\mathcal{F}_5\cosh\mathcal{F}_3\sin\mathcal{F}_6}{\mathcal{F}_2\cosh(\mathcal{F}_3 - \mathcal{F}_4)}\,, &
\mathcal{C}_{19} &= -\frac{\mathcal{F}_5\sinh\mathcal{F}_4\sin\mathcal{F}_6}{\mathcal{F}_1\cosh(\mathcal{F}_3 - \mathcal{F}_4)}\,,\nonumber\\
\mathcal{C}_{11} &= \frac{\mathcal{F}_1\sinh\mathcal{F}_3\cos\mathcal{F}_6}{\mathcal{F}_5}\,, &
\mathcal{C}_{15} &= -\frac{\mathcal{F}_1\sinh\mathcal{F}_3\sin\mathcal{F}_6}{\mathcal{F}_5}\,,\nonumber\\
\mathcal{C}_{12} &= \frac{\mathcal{F}_2\cosh\mathcal{F}_4\cos\mathcal{F}_6}{\mathcal{F}_5}\,, &
\mathcal{C}_{16} &= -\frac{\mathcal{F}_2\cosh\mathcal{F}_4\sin\mathcal{F}_6}{\mathcal{F}_5}\,,\nonumber\\
\mathcal{C}_{13} &= \frac{\mathcal{F}_{5,t}}{\mathcal{F}_5}\,, &
\mathcal{C}_{17} &= \mathcal{F}_{6,t}\,,\nonumber\\
\mathcal{C}_{14} &= \frac{\mathcal{F}_{5,r}}{\mathcal{F}_5}\,, &
\mathcal{C}_{18} &= \mathcal{F}_{6,r}\,.\label{eq:flatparam}
\end{align}
We once again remark that despite being metric-compatible with respect to the metric \(\tilde{g}_{\mu\nu}\), this connection, in general, possesses non-vanishing nonmetricity with respect to the metric \(g_{\mu\nu}\), as we shall see in section~\ref{ssec:norq}. Also its torsion is in general non-vanishing, unless a number of conditions is satisfied, which we discuss in the next section. However, one easily checks that its curvature indeed vanishes.

\subsection{Flat torsion-free: $R = T = 0$}\label{ssec:nort}
We now derive a number of conditions for the Weitzenböck connection of the tetrad~\eqref{eq:sphertetradwb} to be torsion-free. Recall that the torsion of the general spherically symmetric connection is given by the components~\eqref{eq:torsionc}. In particular, we find the condition
\begin{equation}
\mathcal{C}_{20} = 0 \quad \Rightarrow \quad \mathcal{F}_5\cosh\mathcal{F}_3\sin\mathcal{F}_6 = 0\,.
\end{equation}
Here only the last factor may vanish, since for vanishing \(\mathcal{F}_5\) the tetrad~\eqref{eq:sphertetradwb} would be degenerate. Hence, we have \(\mathcal{F}_6/\pi \in \mathbb{Z}\). Without loss of generality, we may set \(\mathcal{F}_6 = 0\). We then continue with the conditions
\begin{equation}
\mathcal{C}_{13} - \mathcal{C}_{11} = \mathcal{C}_{14} - \mathcal{C}_{12} = 0 \quad \Rightarrow \quad \mathcal{F}_{5,t} - \mathcal{F}_1\sinh\mathcal{F}_3 = \mathcal{F}_{5,r} - \mathcal{F}_2\sinh\mathcal{F}_4 = 0\,.
\end{equation}
These conditions determine \(\mathcal{F}_5\) up to a constant of integration, and further impose the integrability condition
\begin{equation}
\partial_r(\mathcal{F}_1\sinh\mathcal{F}_3) = \partial_t(\mathcal{F}_2\cosh\mathcal{F}_4)
\end{equation}
on the remaining parameter functions. These conditions, together with the remaining torsion components \(\mathcal{C}_2 - \mathcal{C}_3 = \mathcal{C}_6 - \mathcal{C}_7 = 0\), finally yield the additional conditions
\begin{subequations}
\begin{align}
\mathcal{F}_{1,r}\cosh(\mathcal{F}_3 - \mathcal{F}_4) + \mathcal{F}_1\mathcal{F}_{3,r}\sinh(\mathcal{F}_3 - \mathcal{F}_4) - \mathcal{F}_2\mathcal{F}_{4,t} &= 0\,,\\
\mathcal{F}_{2,t}\cosh(\mathcal{F}_3 - \mathcal{F}_4) - \mathcal{F}_2\mathcal{F}_{4,t}\sinh(\mathcal{F}_3 - \mathcal{F}_4) - \mathcal{F}_1\mathcal{F}_{3,r} &= 0\,.
\end{align}
\end{subequations}
For any choice of the parameter functions \(\mathcal{F}_3\) and \(\mathcal{F}_4\), one thus obtains a system of coupled, linear, inhomogeneous, first-order, partial differential equations for \(\mathcal{F}_1\) and \(\mathcal{F}_2\). We will not attempt to construct a general solution scheme for these equations, since in general there will be no closed form for the solution.

\subsection{Flat metric-compatible: $R = Q = 0$}\label{ssec:norq}
In order to determine the most general flat and metric-compatible connection, one may proceed similarly to the previously discussed case of a flat and torsion-free connection, imposing that the nonmetricity~\eqref{eq:nonmetc} vanishes. For this purpose, it turns out to be simpler to express the nonmetricity in the components \(g_{tt}, g_{rr}, g_{tr}, g_{\vartheta\vartheta}\) of the metric, without substituting them with the parametrization~\eqref{eq:sphermetric}. In this case we may start with the conditions \(Q_{t\vartheta\vartheta} = Q_{r\vartheta\vartheta} = 0\), which imply
\begin{equation}
\frac{g_{\vartheta\vartheta,t}}{g_{\vartheta\vartheta}} - 2\frac{\mathcal{F}_{5,t}}{\mathcal{F}_5} = \frac{g_{\vartheta\vartheta,r}}{g_{\vartheta\vartheta}} - 2\frac{\mathcal{F}_{5,r}}{\mathcal{F}_5} = 0\,.
\end{equation}
These are solved by \(g_{\vartheta\vartheta} = \mathfrak{g}_1\mathcal{F}_5^2\), where \(\mathfrak{g}_1\) is a constant of integration. To proceed, one may use the purely algebraic equations \(Q_{\varphi t\vartheta} = Q_{\vartheta t\vartheta} = Q_{\varphi r\vartheta} = Q_{\vartheta r\vartheta} = 0\), whose full expression we omit here for brevity. It turns out that these are not independent and may be solved, e.g., for the components \(g_{tt}\) and \(g_{rr}\). The corresponding solution reads
\begin{subequations}\label{eq:gttgrr}
\begin{align}
g_{tt} &= \frac{\mathcal{F}_1}{\mathcal{F}_2\sinh\mathcal{F}_4}\left[g_{tr}\cosh\mathcal{F}_3 - \mathfrak{g}_1\mathcal{F}_1\mathcal{F}_2\sinh\mathcal{F}_3\cosh(\mathcal{F}_3 - \mathcal{F}_4)\right]\,,\\
g_{rr} &= \frac{\mathcal{F}_2}{\mathcal{F}_1\cosh\mathcal{F}_3}\left[g_{tr}\sinh\mathcal{F}_4 + \mathfrak{g}_1\mathcal{F}_1\mathcal{F}_2\cosh\mathcal{F}_4\cosh(\mathcal{F}_3 - \mathcal{F}_4)\right]\,.
\end{align}
\end{subequations}
In order to determine the final component \(g_{tr}\), one uses the remaining components of the nonmetricity~\eqref{eq:nonmetc}. Imposing that these vanish yields a set of first-order partial differential equations for \(g_{tr}\), which are solved by
\begin{equation}
g_{tr} = \mathcal{F}_1\mathcal{F}_2(\mathfrak{g}_1\sinh\mathcal{F}_3\cosh\mathcal{F}_4 - \mathfrak{g}_2\cosh\mathcal{F}_3\sinh\mathcal{F}_4)
\end{equation}
with another constant of integration \(\mathfrak{g}_2\). Finally, substituting this solution into the intermediate result~\eqref{eq:gttgrr} yields the full solution
\begin{equation}
g_{tt} = -\frac{\mathcal{F}_1^2}{2}[\mathfrak{g}_1 + \mathfrak{g}_2 - (\mathfrak{g}_1 - \mathfrak{g}_2)\cosh(2\mathcal{F}_3)]\,, \quad
g_{rr} = \frac{\mathcal{F}_2^2}{2}[\mathfrak{g}_1 + \mathfrak{g}_2 + (\mathfrak{g}_1 - \mathfrak{g}_2)\cosh(2\mathcal{F}_4)]\,.
\end{equation}
We find that the most general flat, metric-compatible metric-affine geometry is determined by the parameter functions \(\mathcal{F}_1, \ldots, \mathcal{F}_6\) and the two constants of integration \(\mathfrak{g}_1\) and \(\mathfrak{g}_2\). Note that for \(\mathfrak{g}_1 = \mathfrak{g}_2 = 1\) the metric \(g_{\mu\nu}\) reduces to the metric \(\tilde{g}_{\mu\nu}\) defined by the tetrad \(\Theta^a{}_{\mu}\), while for general \(\mathfrak{g}_1 = \mathfrak{g}_2\) one obtains a constant multiple of this metric. It is obvious that the connection is compatible with this metric, by construction. It is remarkable, however, that this is not the only solution, and that also metrics with \(\mathfrak{g}_1 \neq \mathfrak{g}_2\) yield metric-compatible geometries.

\subsection{Flat torsion-free metric-compatible: $R = T = Q = 0$}\label{ssec:nortq}
Finally, and again for completeness, we also mention the case in which all three tensorial quantities which characterize the connection - torsion, nonmetricity and curvature - vanish. It is a well-known fact that this condition reduces the metric-affine geometry to Minkowski space, so that the metric is given by the Minkowski metric \(\eta_{\mu\nu}\) and the connection by its Levi-Civita connection. However, this does not become immediately apparent if one calculates the curvature of the Levi-Civita connection~\eqref{eq:levicivita} of the spherically symmetric metric, which implements the conditions of vanishing torsion and nonmetricity, and attempts to solve for vanishing curvature. This is due to the fact that the condition of spherical symmetry commutes with coordinate transformations of the non-angular coordinates \(t, r\), so that the most general spherically symmetric metric in arbitrary, spherical coordinates is given by
\begin{equation}
g_{\mu\nu} = \frac{\partial\tilde{x}^{\rho}}{\partial x^{\mu}}\frac{\partial\tilde{x}^{\sigma}}{\partial x^{\nu}}\eta_{\mu\nu}\,,
\end{equation}
where the new coordinates take the form
\begin{equation}
\tilde{t} = \tilde{t}(t,r)\,, \quad
\tilde{r} = \tilde{r}(t,r)\,, \quad
\tilde{\vartheta} = \vartheta\,, \quad
\tilde{\varphi} = \varphi\,.
\end{equation}
Hence, the components of the most general metric, after imposing spherical symmetry via the conditions~\eqref{eq:metricalgcond}, are given by
\begin{equation}
g_{tt} = \tilde{r}_{,t}^2 - \tilde{t}_{,t}^2\,, \quad
g_{rr} = \tilde{r}_{,r}^2 - \tilde{t}_{,r}^2\,, \quad
g_{tr} = \tilde{r}_{,t}\tilde{r}_{,r} - \tilde{t}_{,t}\tilde{t}_{,r}\,, \quad
g_{\vartheta\vartheta} = \tilde{r}^2\,.
\end{equation}
One easily checks that the curvature of the Levi-Civita connection induced by this metric indeed vanishes.

\section{Reflection symmetry}\label{sec:reflect}
During the preceding sections, we have understood as the rotation group only the proper rotations, which form the connected Lie group \(\mathrm{SO}(3)\). Only for this connected group the infinitesimal treatment we used is equivalent to a treatment using finite group actions, as discussed in section~\ref{sec:symmetry}. One may easily extend this analysis to the (general) orthogonal group \(\mathrm{O}(3)\) by also allowing for reflections. Since this group consists of two connected components, one must use the transformation laws~\eqref{eq:metsymcondf} and~\eqref{eq:affsymcondf} of the metric and affine connection under finite group actions for at least one reflection. It is convenient to consider the equatorial reflection\footnote{Note that it might seem a more straightforward and canonical choice to consider the point reflection, which in addition also replaces \(\varphi\) by \(\varphi' = \varphi + \pi\). However, since the latter is simply a rotation around the polar axis, which is already covered by the symmetry under proper rotations, it is equivalent to the choice we make here.}
\begin{equation}
x \mapsto x' = \phi(x)\,, \quad
(t, r, \vartheta, \varphi) \mapsto (t', r', \vartheta', \varphi') = (t, r, \pi - \vartheta, \varphi)\,.
\end{equation}
Together with the Jacobian, whose non-vanishing components are given by
\begin{equation}
\frac{\partial t'}{\partial t} = \frac{\partial r'}{\partial r} = \frac{\partial\varphi'}{\partial\varphi} = 1\,, \quad
\frac{\partial\vartheta'}{\partial\vartheta} = -1\,,
\end{equation}
we find the following transformation rules for the metric and connection coefficients:
\begin{enumerate}
\item
If the number of coordinate indices \(\vartheta\) on \(g_{\mu\nu}\) or \(\Gamma^{\mu}{}_{\nu\rho}\) is odd, a factor \(-1\) is incurred.
\item
Due to the coordinate change \(\vartheta \mapsto \pi - \vartheta\), all occurrences of \(\cos\vartheta\) are replaced by \(-\cos\vartheta\), while \(\sin\vartheta\) is retained. This also propagates to constructed triangular functions such as \(\cot\vartheta\).
\end{enumerate}
With these transformation rules in mind, one finds that the most general spherically symmetric metric~\eqref{eq:sphermetric} is also invariant under reflections. Hence, enlarging the symmetry group from \(\mathrm{SO}(3)\) to \(\mathrm{O}(3)\) does not impose any constraints on the parameter functions \(\mathcal{G}_1, \ldots, \mathcal{G}_4\). However, the situation is different for the affine connection. Here imposing invariance of the connection coefficients~\eqref{eq:sphergamma} under reflections imposes the additional constraints
\begin{equation}\label{eq:reflcondc}
\mathcal{C}_{15} = \mathcal{C}_{16} = \mathcal{C}_{17} = \mathcal{C}_{18} = \mathcal{C}_{19} = \mathcal{C}_{20} = 0\,.
\end{equation}
Only connections whose parameter functions satisfy these constraints are also symmetric under \(\mathcal{O}(3)\).

The same discussion as shown above for the metric and affine connection can also be carried over to the formulation in terms of a tetrad and spin connection as shown in section~\ref{ssec:tetsym}. Applying the appropriate symmetry conditions to the tetrad and spin connection displayed in section~\ref{ssec:tetspin}, one finds that these are invariant under reflections if and only if the conditions
\begin{equation}\label{eq:reflconds}
\mathcal{S}_{15} = \mathcal{S}_{16} = \mathcal{S}_{17} = \mathcal{S}_{18} = \mathcal{S}_{19} = \mathcal{S}_{20} = 0
\end{equation}
are imposed, in full analogy to the metric-Palatini formulation.

It is also instructive to study how the additional conditions imposed by reflection symmetry affect the tensorial properties of the metric-affine geometry, which we discussed in section~\ref{sec:props}. For this purpose it is helpful to first rewrite the conditions in terms of the parameter functions \(\mathcal{T}_1, \ldots, \mathcal{T}_8, \mathcal{Q}_1, \ldots, \mathcal{Q}_{12}\) introduced in section~\ref{ssec:conndecomp}, which yields
\begin{equation}\label{eq:reflcondtq}
\mathcal{T}_3 = \mathcal{T}_4 = \mathcal{T}_7 = \mathcal{T}_8 = \mathcal{Q}_{11} = \mathcal{Q}_{12} = 0\,.
\end{equation}
This means that only 4 of the formerly 8 independent components of the torsion are left, while 10 of the formerly 12 independent components of the nonmetricity are left. One easily checks which components of the torsion~\eqref{eq:torsiont}, nonmetricity~\eqref{eq:nonmetq}, contortion~\eqref{eq:contortiont} and disformation~\eqref{eq:disformationq} vanish as a consequence of these conditions, and how the non-vanishing components are parametrized by the remaining parameter functions.

A particular amount of simplifications is also obtained for the curvature, which we discussed in section~\ref{ssec:curv}. For the components~\eqref{eq:curv1} we find that all except for the last line vanish if reflection symmetry is imposed, while the latter simplifies to
\begin{equation}\label{eq:curvsimp}
R^{\vartheta}{}_{\varphi\vartheta\varphi} = -R^{\varphi}{}_{\vartheta\vartheta\varphi}\sin^2\vartheta = \left(1 + \mathcal{C}_9\mathcal{C}_{11} + \mathcal{C}_{10}\mathcal{C}_{12}\right)\sin^2\vartheta\,.
\end{equation}
Similarly, the second line of the components~\eqref{eq:curv2} vanishes under reflection symmetry, while its first line as well as the components~\eqref{eq:curv3} remain unchanged. Finally, the second half of the components~\eqref{eq:curv4} also vanishes if reflection symmetry is imposed, while the first half slightly simplifies, similarly to the component~\eqref{eq:curvsimp}.

The restrictions imposed on the affine connection and its curvature can also be studied for the more specific classes of metric-affine geometries discussed in section~\ref{sec:special}. We first remark that for the torsion-free and metric-compatible geometries shown in sections~\ref{ssec:not} and~\ref{ssec:noq} one immediately obtains the reflection invariant subclasses by imposing the conditions~\eqref{eq:reflcondtq} on the parameter functions. For the Levi-Civita connection shown in section~\ref{ssec:notq}, reflection symmetry automatically follows from the reflection symmetry of the metric~\eqref{eq:sphermetric}, so that no additional constraints are obtained in this case. In the flat (curvature-free) case discussed in section~\ref{ssec:nor}, one finds that reflection symmetry is imposed by the condition\footnote{More precisely, the condition obtained is \(\mathcal{F}_6/\pi \in \mathbb{Z}\), but one may set \(\mathcal{F}_6 = 0\), as mentioned in section~\ref{ssec:nort}.} \(\mathcal{F}_6 = 0\), while the remaining parameter functions \(\mathcal{F}_1, \ldots, \mathcal{F}_5\) remain unconstrained. Note that this condition is always satisfied in the flat, torsion-free case studied in section~\ref{ssec:nort}, while it must be imposed as an independent constraint in the flat, metric-compatible case shown in section~\ref{ssec:norq}. Of course, the Minkowski metric obtained in section~\ref{ssec:nortq} is invariant under reflections.

\section{Autoparallel motion}\label{sec:autoparallel}
As an example for a potential physical application of our findings, we study the orbital motion of a hypothetical class of test particles which follow the autoparallels of the affine connection. Such kind of motion may arise from considering a generalized coupling of test matter to gravity, such as a generalized fluid, or generalized observer frames~\cite{Manoff:1999ae,Manoff:2000xq}. Hence, we will consider trajectories \(\gamma: \mathbb{R} \to M\) which are subject to the autoparallel equation
\begin{equation}\label{eq:autoparallel}
\ddot{\gamma}^{\rho} + \Gamma^{\rho}{}_{\mu\nu}\dot{\gamma}^{\mu}\dot{\gamma}^{\nu} = 0\,,
\end{equation}
where \(\Gamma^{\rho}{}_{\mu\nu}\) are the coefficients of the most general spherically symmetric connection we derived in section~\ref{ssec:conn}. To further simplify the task at hand, we restrict ourselves to stationary metric-affine geometries, so that the functions parametrizing the connection depend only on \(r\) and not on \(t\). Further, we will restrict ourselves to circular orbits parallel to the equatorial plane, which are of the form
\begin{equation}\label{eq:circautop}
\gamma^a(\tau) = (N\tau, R, \Theta, \Omega\tau)
\end{equation}
with constant parameters \(N, R, \Theta, \Omega\). It is well known that in Riemannian geometry, where the connection coefficients are given by the Levi-Civita connection, such orbits are necessarily coplanar with the center of spherical symmetry, hence \(\Theta = \frac{\pi}{2}\), i.e., they lie in the equatorial plane, due to the conservation of angular momentum. This follows from the spherical symmetry of the metric background geometry, which imposes the same symmetry on the test body Lagrangian, in conjunction with Noether's first theorem. However, this line of argument does not hold for general autoparallel motion, which is not necessarily derived from a Lagrangian.

Inserting the ansatz~\eqref{eq:circautop} into the autoparallel equation~\eqref{eq:autoparallel}, we find the component equations
\begin{subequations}\label{eq:autocomp}
\begin{align}
\mathcal{C}_1N^2 + \mathcal{C}_9\Omega^2\sin^2\Theta &= 0\,,\label{eq:autot}\\
\mathcal{C}_5N^2 + \mathcal{C}_{10}\Omega^2\sin^2\Theta &= 0\label{eq:autor}\,,\\
\Omega\cos\Theta + (\mathcal{C}_{15} + \mathcal{C}_{17}) &= 0\,,\label{eq:autotheta}\\
\mathcal{C}_{11} + \mathcal{C}_{13} &= 0\,.\label{eq:autophi}
\end{align}
\end{subequations}
These equations can be interpreted as follows:
\begin{enumerate}
\item
Equation~\eqref{eq:autot} imposes the constancy of the lapse parameter \(N\). This equation could be absorbed by considering a more general parametrization of the trajectory, or a transformation of the time coordinate \(t \mapsto t'\).
\item
Equation~\eqref{eq:autor} imposes the constancy of the radial coordinate \(R\). Its left hand side can be interpreted as the sum of radial gravitational and fictuous forces along the orbit, which vanishes once the orbit satisfies the autoparallel equation~\eqref{eq:autoparallel}.
\item
Similarly to the previous one, equation~\eqref{eq:autotheta} imposes the constancy of the azimuth angle \(\Theta\). The left hand side of this equation represents a force which is tangent to the sphere of radius \(R\) and perpendicular to the trajectory \(\gamma\). It is in particular notable that such a force occurs also for motion in the equatorial plane, \(\Theta = \frac{\pi}{2}\), in case the connection coefficients satisfy \(\mathcal{C}_{15} + \mathcal{C}_{17} \neq 0\). In this case it follows that there are no autoparallel circular orbits in the equatorial plane. Also note that the resulting perpendicular force acting on this orbit breaks reflection symmetry, since it imposes a preferred orientation. This is consistent with our findings from section~\ref{sec:reflect}, that reflection symmetry imposes \(\mathcal{C}_{15} = \mathcal{C}_{17} = 0\)\,.
\item
Finally, equation~\eqref{eq:autophi} imposes the constancy of the angular frequency \(\Omega\). The corresponding term \(\mathcal{C}_{11} + \mathcal{C}_{13}\) can be interpreted as a longitudinal force along the trajectory, leading to an acceleration with respect to the chosen parametrization.
\end{enumerate}
Observe in particular that the equations~\eqref{eq:autocomp} do not admit a simultaneous solution unless there exists a constant radius \(R\) at which the (radial coordinate dependent) connection coefficients satisfy \(\mathcal{C}_{11} + \mathcal{C}_{13} = 0\) and \(\mathcal{C}_1\mathcal{C}_{10} - \mathcal{C}_5\mathcal{C}_9 = 0\). If these conditions are satisfied, one can solve the autoparallel equation for the remaining parameters \(N, \Theta, \Omega\).

It is also instructive to view this result in the light of the decomposition of the connection which we presented in section~\ref{ssec:conndecomp}. We observe the following:
\begin{enumerate}
\item
The coupled equations~\eqref{eq:autot} and~\eqref{eq:autor} are jointly influenced by the torsion components \(\mathcal{T}_1, \mathcal{T}_2, \mathcal{T}_5, \mathcal{T}_6\) and nonmetricity components \(\mathcal{Q}_1, \mathcal{Q}_2, \mathcal{Q}_3, \mathcal{Q}_4, \mathcal{Q}_5, \mathcal{Q}_8, \mathcal{Q}_9, \mathcal{Q}_{10}\). These components affect both the lapse and the radial force along the studied orbit.
\item
The transversal force equation~\eqref{eq:autotheta} depends on the components \(\mathcal{T}_3, \mathcal{T}_4, \mathcal{T}_7\) of the torsion and \(\mathcal{Q}_{11}\) of the nonmetricity, which vanish if reflection symmetry is imposed.
\item
The last equation~\eqref{eq:autophi} depends on the torsion and nonmetricity components \(\mathcal{T}_5\) and \(\mathcal{Q}_4\), so that we can associate these with a longitudinal force along the orbit we studied.
\end{enumerate}
In summary, we find that coupling test matter to torsion or nonmetricity influences both the orbital parameters and possible existence of circular orbits, including the possibility of shifting the orbit out of the equatorial plane if reflection symmetry breaking terms are present.

\section{Cosmological symmetry}\label{sec:cosmo}
The results we derived in the previous sections of this article can easily be extended from spherical to cosmological symmetry. For this purpose one introduces the additional symmetry generating vector fields
\begin{subequations}\label{eq:genvectrans}
\begin{align}
X_1 &= \chi\sin\vartheta\cos\varphi\partial_r + \frac{\chi}{r}\cos\vartheta\cos\varphi\partial_{\vartheta} - \frac{\chi\sin\varphi}{r\sin\vartheta}\partial_{\varphi}\,,\\
X_2 &= \chi\sin\vartheta\sin\varphi\partial_r + \frac{\chi}{r}\cos\vartheta\sin\varphi\partial_{\vartheta} + \frac{\chi\cos\varphi}{r\sin\vartheta}\partial_{\varphi}\,,\\
X_3 &= \chi\cos\vartheta\partial_r - \frac{\chi}{r}\sin\vartheta\partial_{\vartheta}\,,
\end{align}
\end{subequations}
where we made use of the abbreviation \(\chi = \sqrt{1 - kr^2}\), and where \(k \in \{-1,0,1\}\) is the sign of the spatial curvature. Using the spherically symmetric metric-affine geometry we derived, it turns out to be sufficient to impose symmetry under the last generator \(X_3\), since symmetry under the remaining generators then follows from their commutation relations. For the metric this yields the well-known Robertson-Walker metric
\begin{equation}
g_{tt} = -\mathcal{N}^2\,, \quad
g_{rr} = \frac{\mathcal{A}^2}{1 - kr^2}\,, \quad
g_{\vartheta\vartheta} = \mathcal{A}^2r^2\,, \quad
g_{\varphi\varphi} = g_{\vartheta\vartheta}\sin^2\vartheta\,,
\end{equation}
which is parametrized by the lapse \(\mathcal{N}(t)\) and scale factor \(\mathcal{A}(t)\). For the connection one obtains the algebraic equations
\begin{gather}
\mathcal{C}_2 = \mathcal{C}_3 = \mathcal{C}_5 = \mathcal{C}_{15} = \mathcal{C}_{17} = \mathcal{C}_{19} = 0\,, \quad \mathcal{C}_{12} = \mathcal{C}_{14} = \frac{1}{r}\,, \quad \mathcal{C}_{10} = r(kr^2 - 1)\,,\\
\mathcal{C}_8 = \frac{kr}{1 - kr^2}\,, \quad \mathcal{C}_6 = \mathcal{C}_{11}\,, \quad \mathcal{C}_7 = \mathcal{C}_{13}\,, \quad \mathcal{C}_4 = \frac{\mathcal{C}_9}{r^2(kr^2 - 1)}\,, \quad \mathcal{C}_{16} = -\mathcal{C}_{18} = \frac{\mathcal{C}_{20}}{r^2(kr^2 - 1)}\,,\nonumber
\end{gather}
which determine 15 components of \(\mathcal{C}\) in terms of the remaining 5 components. These remaining independent components are constrained by the differential equations
\begin{equation}
\partial_r\mathcal{C}_1 = \partial_r\mathcal{C}_{11} = \partial_r\mathcal{C}_{13} = \partial_r\mathcal{C}_9 - 2\frac{\mathcal{C}_9}{r} = \partial_r\mathcal{C}_{20} - \frac{2 - 3kr^2}{r(1 - kr^2)}\mathcal{C}_{20} = 0\,.
\end{equation}
The most general solution to these equations depends on 5 functions \(\mathcal{K}_1(t), \ldots, \mathcal{K}_5(t)\). In terms of these, the parameter functions of the most general spherically symmetric connection are expressed as
\begin{gather}
\mathcal{C}_1 = \mathcal{K}_1\,, \quad \mathcal{C}_6 = \mathcal{C}_{11} = \mathcal{K}_3\,, \quad \mathcal{C}_7 = \mathcal{C}_{13} = \mathcal{K}_4\,, \quad \mathcal{C}_4 = \frac{\mathcal{K}_2}{1 - kr^2}\,, \quad \mathcal{C}_9 = \mathcal{K}_2r^2\,, \quad \mathcal{C}_8 = \frac{kr}{1 - kr^2}\,,\nonumber\\
\mathcal{C}_{20} = \mathcal{K}_5r^2\sqrt{1 - kr^2}\,, \quad \mathcal{C}_{18} = -\mathcal{C}_{16} = \frac{\mathcal{K}_5}{\sqrt{1 - kr^2}}\,, \quad \mathcal{C}_{12} = \mathcal{C}_{14} = \frac{1}{r}\,, \quad \mathcal{C}_{10} = r(kr^2 - 1)\,.
\end{gather}
Inserting these parameter functions in the general spherically symmetric connection~\eqref{eq:sphergamma} finally yields its explicit form
\begin{gather}
\Gamma^t{}_{tt} = \mathcal{K}_1\,, \quad \Gamma^r{}_{tr} = \Gamma^{\vartheta}{}_{t\vartheta} = \Gamma^{\varphi}{}_{t\varphi} = \mathcal{K}_3\,, \quad \Gamma^r{}_{rt} = \Gamma^{\vartheta}{}_{\vartheta t} = \Gamma^{\varphi}{}_{\varphi t} = \mathcal{K}_4\,, \quad \Gamma^t{}_{rr} = \frac{\mathcal{K}_2}{1 - kr^2}\,,\nonumber\\
\Gamma^t{}_{\vartheta\vartheta} = \mathcal{K}_2r^2\,, \quad \Gamma^t{}_{\varphi\varphi} = \mathcal{K}_2r^2\sin^2\vartheta\,, \quad \Gamma^r{}_{\varphi\vartheta} = -\Gamma^r{}_{\vartheta\varphi} = \mathcal{K}_5r^2\sqrt{1 - kr^2}\sin\vartheta\,,\\
\Gamma^{\vartheta}{}_{r\varphi} = -\Gamma^{\vartheta}{}_{\varphi r} = \frac{\mathcal{K}_5\sin\vartheta}{\sqrt{1 - kr^2}}\,, \quad \Gamma^{\varphi}{}_{r\vartheta} = -\Gamma^{\varphi}{}_{\vartheta r} = -\frac{\mathcal{K}_5}{\sqrt{1 - kr^2}\sin\vartheta}\,, \quad \Gamma^r{}_{rr} = \frac{kr}{1 - kr^2}\,,\nonumber\\
\Gamma^{\vartheta}{}_{r\vartheta} = \Gamma^{\vartheta}{}_{\vartheta r} = \Gamma^{\varphi}{}_{r\varphi} = \Gamma^{\varphi}{}_{\varphi r} = \frac{1}{r}\,, \quad \Gamma^{\varphi}{}_{\vartheta\varphi} = \Gamma^{\varphi}{}_{\varphi\vartheta} = \cot\vartheta\,, \quad \Gamma^{\vartheta}{}_{\varphi\varphi} = -\sin\vartheta\cos\vartheta\,,\nonumber\\
\Gamma^r{}_{\vartheta\vartheta} = r(kr^2 - 1)\,, \quad \Gamma^r{}_{\varphi\varphi} = r(kr^2 - 1)\sin^2\vartheta\,.\nonumber
\end{gather}
If one also imposes reflection symmetry, as discussed in section~\ref{sec:reflect}, one obtains the additional condition \(\mathcal{K}_5 = 0\). One can now perform the same kind of analysis as in the case of spherical symmetry shown in the previous sections. We will not perform such calculations here, as they would exceed the scope of this article, whose aim is the discussion of the spherically symmetric case, and refer to~\cite{Minkevich:1998cv}, where such kind of analysis for cosmology is performed in a different parametrization.

\section{Conclusion}\label{sec:conclusion}
We showed how to construct the most general metric-affine geometry with spherical symmetry and studied its properties. We demonstrated that it is determined by 4 parameter functions which determine the metric, as well as 20 parameter functions which determine the connection. We further decomposed the latter into 8 components determining the torsion and 12 components determining the nonmetricity. This decomposition allowed us to derive a simple parametrization for those metric-affine geometries where either of these two tensorial quantities vanishes, similarly to the parametrization found in~\cite{Minkevich:2003it}. Furthermore, we calculated the curvature, and constructed the most general flat metric-affine geometries with spherical symmetry, whose connection is determined by 6 parameter functions. Finally, we gave conditions on the torsion-free case and determined the most general metric-affine geometry with vanishing nonmetricity. As an interesting result, we found a two-parameter family of metrics which are compatible with the most general flat connection.

By extending the symmetry group from the proper rotations \(\mathrm{SO}(3)\) to the full orthogonal group \(\mathrm{O}(3)\), we found that the metric automatically satisfies also this larger symmetry, so that no additional restrictions on its parameter functions are obtained, while we found 6 further constraints on the coefficients of the affine connection, hence leaving only 14 free functions parametrizing the affine connection, of which 4 determine the torsion, while 10 determine the nonmetricity. In the case of a flat affine connection, we obtained one condition on the parameter functions imposed by reflection symmetry, leaving 5 free functions parametrizing the connection coefficients. In particular, we saw that this condition is always satisfied if the connection is not only flat, but also torsion-free.

To demonstrate a possible physical application of our result, we studied circular orbits for a hypothetical class of test bodies which follow the autoparallels of the affine connection, and showed that circular orbits may not exist in general. Also, in contrast to purely Riemannian geometry with spherical symmetry, we found the possibility of circular orbits which are not coplanar with the center or spherical symmetry, but shifted with respect to the equatorial plane.

Given any gravitational theory based on the metric-affine geometry or one of its subclasses, defined by the vanishing of torsion, nonmetricity or curvature, one may use the corresponding spherically symmetric geometry as an ansatz to solve the field equations. The explicit expressions in different parametrizations, which are adapted to the particular subclass and which we provide in this article, may serve as utilities in this task. Hence, the work presented in this article should be seen as a basic ingredient towards the classification of spherically symmetric solutions to gravity theories based on metric-affine geometries; the second ingredient, which is the choice of such gravity theories and hence their field equations, finding the solutions and their physical properties, is left for future work.

Various modifications and generalizations of the calculations shown here are possible. Instead of spherical symmetry, one may consider, e.g., planar symmetry to study exact planar wave solutions, or cosmological symmetry. For the latter, we have provided the most general metric-affine geometry as well, which can serve as a starting point for such kind on calculations, similarly to the work presented in~\cite{Minkevich:1998cv}. Another possibility is to consider other types of geometries based on Cartan geometry, which are relevant in physics and to which the notion of spacetime symmetries derived from Cartan geometry~\cite{Hohmann:2015pva} applies. Possible generalizations include bimetric geometries, where different branches exist depending on whether the two metrics can be simultaneously brought to diagonal form or not~\cite{Comelli:2011wq}, as well as Finsler geometries, where the geometry is defined on the tangent bundle instead of the spacetime manifold itself, and where spherical symmetry can be implemented via the action of the rotation group on distinguished tensor fields on the tangent bundle~\cite{Hohmann:2018hac}. Also the connection between spacetime symmetries and conservation laws~\cite{Cattani:1993} may be explored using the methods we discussed.

\vspace{6pt}




\funding{This work was supported by the Estonian Research Council through the Personal Research Funding project PRG356 and by the European Regional Development Fund through the Center of Excellence TK133 ``The Dark Side of the Universe''.}

\acknowledgments{The author thanks S. Capozziello, O. Luongo and R. Giambò for the kind invitation to contribute to this Special Issue.}

\conflictsofinterest{The author declares no conflict of interest.}

%

%

\reftitle{References}


\externalbibliography{yes}
\bibliographystyle{unsrt}
\bibliography{spher}

\end{document}